\documentclass[journal,10pt]{IEEEtran}
\usepackage{amssymb}
\usepackage{amsmath}
\usepackage{cite}
\usepackage{url}
\usepackage{xcolor}
\usepackage{cite,graphicx,amsmath,amssymb}
\usepackage{subfigure}
\usepackage{fancyhdr}
\usepackage{mdwmath}
\usepackage{mdwtab}
\usepackage{caption}
\usepackage{amsthm}
\usepackage{setspace}
\usepackage{hyperref}
\usepackage{algorithm}
\usepackage{algorithmic}
\hypersetup{colorlinks=false}
\usepackage{array}
\usepackage{makecell}
\usepackage{threeparttable}
\usepackage{tabularx,colortbl}
\usepackage{multirow, booktabs}
\usepackage{diagbox}

\newcommand{\bm}[1]{\mbox{\boldmath{$#1$}}}

\newtheorem{remark}{Remark}
\newtheorem{theorem}{Theorem}

\newtheorem{lemma}{Lemma}

\newtheorem{corollary}{Corollary}

\allowdisplaybreaks
\makeatletter
\def\ScaleIfNeeded{%
\ifdim\Gin@nat@width>\linewidth \linewidth \else \Gin@nat@width
\fi } \makeatother

\begin{document}
\title{NOMA-Aided Joint Radar and Multicast-Unicast Communication Systems}
\author{

Xidong~Mu,~\IEEEmembership{Graduate Student Member,~IEEE,}
        Yuanwei~Liu,~\IEEEmembership{Senior Member,~IEEE,}
       Li~Guo,~\IEEEmembership{Member,~IEEE,}
       Jiaru~Lin,~\IEEEmembership{Member,~IEEE,}
       and Lajos~Hanzo,~\IEEEmembership{Fellow,~IEEE}

\thanks{Part of this work will be presented at the IEEE International Conference on Communications (ICC), Seoul, South Korea, May 16-20, 2022~\cite{Mu2022}.}
\thanks{Xidong Mu, Li Guo, and Jiaru Lin are with the Key Laboratory of Universal Wireless Communications, Ministry of Education, Beijing University of Posts and Telecommunications, Beijing 100876, China, and are also with the School of Artificial Intelligence, Beijing University of Posts and Telecommunications, Beijing 100876, China. (e-mail: muxidong@bupt.edu.cn; guoli@bupt.edu.cn; jrlin@bupt.edu.cn).}
\thanks{Yuanwei Liu is with the School of Electronic Engineering and Computer Science, Queen Mary University of London, London E1 4NS, U.K. (email:yuanwei.liu@qmul.ac.uk).}
\thanks{Lajos Hanzo is with the School of Electronics and Computer Science, the University of Southampton, Southampton SO17 1BJ, U.K. (e-mail: lh@ecs.soton.ac.uk).}
}

\maketitle
\vspace{-1.3cm}
\begin{abstract}
The novel concept of non-orthogonal multiple access (NOMA) aided joint radar and multicast-unicast communication (Rad-MU-Com) is investigated. Employing the same spectrum resource, a multi-input-multi-output (MIMO) dual-functional radar-communication (DFRC) base station detects the radar-centric users (R-user), while transmitting mixed multicast-unicast messages both to the R-user and to the communication-centric user (C-user). In particular, the multicast information is intended for both the R- and C-users, whereas the unicast information is only intended for the C-user. More explicitly, NOMA is employed to facilitate this \emph{double spectrum sharing}, where the multicast and unicast signals are superimposed in the power domain and the superimposed communication signals are also exploited as radar probing waveforms. First, a \emph{beamformer-based} NOMA-aided joint Rad-MU-Com framework is proposed for the system having a single R-user and a single C-user. Based on this framework, the unicast rate maximization problem is formulated by optimizing the beamformers employed, while satisfying the rate requirement of multicast and the predefined accuracy of the radar beam pattern. The resultant non-convex optimization problem is solved by a penalty-based iterative algorithm to find a high-quality near-optimal solution. Next, the system is extended to the scenario of multiple pairs of R- and C-users, where a \emph{cluster-based} NOMA-aided joint Rad-MU-Com framework is proposed. A joint beamformer design and power allocation optimization problem is formulated for the maximization of the sum of the unicast rate at each C-user, subject to the constraints on both the minimum multicast rate for each R\&C pair and on accuracy of the radar beam pattern for detecting multiple R-users. The resultant joint optimization problem is efficiently solved by another penalty-based iterative algorithm developed. Finally, our numerical results reveal that significant performance gains can be achieved by the proposed schemes over the benchmark schemes employing conventional transmission strategies.
\end{abstract}

\section{Introduction}
Given the rapid development of cost-efficient electronic technologies, the number of connected devices (e.g., smart phones and Internet-of-Things (IoT) nodes) in the wireless networks escalates. It is forecast that the global mobile data traffic in 2022 will be seven times of that in 2017~\cite{Cisco}. Moreover, new attractive applications (e.g., virtual reality (VR), augmented reality (AR), and ultra-high definition (UHD) video streaming) have emerged, which significantly improve the user-experience, but exacerbate the spectral congestion. As a remedy, a promising solution is to harness spectrum sharing between radar and communication systems~\cite{survey}.\\
\indent Radar (which is short for ``radio detection and ranging'') was originally proposed for military applications in the 1930s, and has rapidly developed in the past decades for both civilian and military applications~\cite{radar}. In contrast to wireless communications, where the radio waves convey information bits, radar employs radio waves to determine the target's characteristics (e.g., location, velocity, shape, etc.) by first transmitting probing signals and then analyzing the received echoes reflected by the target. The superior and necessity of carrying out spectrum sharing between radar and communication are summarized as follows:
\begin{itemize}
  \item On the one hand, radar systems occupy a large amount of spectrum for both civilian and military applications~\cite{survey,survey2}. For instance, spectrum bands below 10 GHz, such as the S-band (2-4 GHz) and C-band (4-8 GHz), are occupied by radar systems used for weather observation. Additionally, the mmWave bands (30-300 GHz), which recently attracted extensive research interests in 5G networks, have already been used by radar systems for collision avoidance in autonomous driving and high-resolution imaging. Therefore, spectrum sharing with radar would allow communication systems to glean additional spectrum resources.
  \item On the other hand, the integration of radar and communication would support promising but challenging near-future applications. For instance, simultaneously supporting both radar and communication functions is essential for autonomous vehicles (AVs)~\cite{AV}, where the driving safety can be guaranteed via the real-time collision avoidance provided by radar and the reliable information exchange between AVs and their controllers.
\end{itemize}
\subsection{State-of-the-art}
In recent years, there have been growing research interests in communication and radar spectrum sharing (CRSS). Generally speaking, the existing research contributions may be classified into two categories: (1) radar and communication coexistence (RCC)~\cite{RCC} and (2) dual-function radar communication (DFRC)~\cite{DFRC}. The key difference between the two categories is whether the radar and communication functions are facilitated by two separate systems or a common system.
\subsubsection{Studies on RCC} The goal of RCC is to efficiently manage the mutual interference between the radar and communication systems, so that they can use the same spectrum to accomplish their own tasks. Saruthirathanaworakun {\em et al.}~\cite{Opportunistic} proposed an opportunistic spectrum sharing scheme, where the communication system can occasionally deliver its information, when the radar spectrum is idle. Exploiting multiple antennas, Sodagari {\em et al.}~\cite{null} designed a multi-input-multi-output (MIMO) radar beamformer (BF) with the objective of projecting the radar probing signals into the null space of the interference channel between the radar and communication systems. Moreover, Li {\em et al.}~\cite{Bo} developed a cooperative spectrum sharing scheme for RCC, where the communication system's transmit covariance matrix and the radar sampling scheme were jointly designed for minimizing the radar's received interference power. Upon relying on realistic imperfect CSI, Liu {\em et al.}~\cite{robust} conceived a robust BF design for RCC to improve the radar's detection performance, while satisfying the communication requirements. Qian {\em et al.}~\cite{Qian} jointly optimized the radar and communication systems for maximizing either the radar's received signal-to-interference plus noise ratio (SINR) or the communication rate achieved. In contrast to the above contributions only aiming for mitigating the interference, Liu {\em et al.}~\cite{Fan_Exploitation} proposed a novel symbol-level precoding scheme for RCC capable of constructively leveraging the multiuser interference. D'Andrea {\em et al.}~\cite{Carmen} studied the effect of a wide-beam search based radar on the uplink performance of a massive MIMO communication system. As a further advance, Wang {\em et al.}~\cite{Jingjing} invoked machine learning tools for beneficially configuring the network association scheme for the communication user in RCC, where the communication throughput was maximized, while simultaneously coordinating the radar's received interference.
\subsubsection{Studies on DFRC} As the functions of radar and communication are facilitated using a joint platform, the resultant hardware cost of DFRC is significantly reduced compared to RCC. Hence, DRFC has recently become a focal point of the CRSS research field. Hassanien {\em et al.}~\cite{sidelobe} exploited the sidelobe of the radar beam to embed information bits into it for communication users. Liu {\em et al.}~\cite{protocol} proposed a pair of sophisticated strategies for implementing DFRC, namely a separated and a shared deployment, with the aim of constructing a high-quality radar beam pattern, while satisfying the communication requirements. As a further development, based on the separated and shared deployment, Dong {\em et al.}~\cite{low} and Liu {\em et al.}~\cite{radio} conceived low-complexity BF design algorithms, respectively. As an innovative contribution, Liu {\em et al.}~\cite{optimal} studied the optimal waveform design of DFRC under the shared deployment paradigm, where branch-and-bound based algorithms were developed. Furthermore, Wang {\em et al.}~\cite{Sparse} studied the employment of sparse arrays for DFRC, where three schemes were proposed for improving the communication performance in the presence of radar detection. Moreover, Huang {\em et al.}~\cite{index} and Ma {\em et al.}~\cite{SM} employed index modulation and spatial modulation techniques for DFRC, respectively. Su {\em et al.}~\cite{security} studied the secure transmission for DFRC, where the radar target was treated as an eavesdropper and artificial noise was employed for degrading its received SINR, while satisfying the communication requirements of the legitimate users. To reveal the fundamental performance limits of DFRC, Chen {\em et al.}~\cite{Pareto} proposed a Pareto optimization framework, where a performance region was defined relying on the difference between the peak and sidelobe (DPSL) of the radar and the SINR achieved by the communication user. Hua {\em et al.}~\cite{Optimal_Jie} proposed two types of receiver structures for the DFRC system in terms of whether the sensing interference can be eliminated or not. Based on the two structures, the optimal BF for minimizing the radar beam pattern error was derived.
\subsection{Motivations and Contributions}
On the one hand, despite the fact that CRSS allows the communication systems to occupy more spectral resources, it is still essential to further improve the spectral efficiency (SE) for satisfying the stringent communication requirements of next-generation wireless networks. In this context, non-orthogonal multiple access (NOMA) is regarded as a promising technique of improving the communication performance, which also shares the idea of ``spectrum sharing''~\cite{Liu2017,Ding}. By employing superposition coding (SC) and successive interference cancellation (SIC) at the transmitters and receivers, respectively, NOMA allows multiple users to share the same frequency resources and distinguishes them in the power domain. Compared to orthogonal multiple access (OMA), NOMA can benefit CRSS by serving more users and hence achieving higher SE. On the other hand, note that the aforementioned research contributions on CRSS only considered unicast communications and assumed that the radar target does not communicate, it merely has to be detected. This paradigm represents a pair of \emph{isolated} systems. Given the diverse future applications of wireless networks, more sophisticated CRSS schemes have to be conceived for supporting mixed multicast-unicast communication and simultaneously communicating with and detecting the radar target user. To this end, the employment of NOMA can provide flexible resource allocation and information transmission options for CRSS. Nevertheless, to the best of our knowledge, the interplay between NOMA as well as CRSS and the potential performance gain have not been studied, which provides the main motivation of this work.\\
\begin{table*}[!ht]\large
\caption{Our contributions in contrast to the state-of-the-art.}
\begin{center}
\centering
\resizebox{\textwidth}{!}{
\begin{tabular}{!{\vrule width1.4pt}l!{\vrule width1.4pt}c!{\vrule width1.4pt}c!{\vrule width1.4pt}c!{\vrule width1.4pt}c!{\vrule width1.4pt}c!{\vrule width1.4pt}c!{\vrule width1.4pt}}
\Xhline{1.4pt}
\centering
& \cite{sidelobe} &\cite{protocol} & \cite{low,radio}&\cite{optimal}&\cite{security}& \bf{Proposed} \\
\Xhline{1.4pt}
\centering
Single Radar target & $\surd$ & $\surd$ & $\surd$, but not mentioned  &$\surd$ &$\surd$ &$\surd$ \\
\hline
\centering
Multiple Radar targets & $\times$ & $\surd$ & $\surd$  &$\surd$ &$\times$ &$\surd$ \\
\hline
\centering
Radar target communication requirement & $\times$ & $\times$ & $\times$ &$\times$ &$\times$ &$\surd$ \\
\hline
\centering
Single communication user & $\surd$ & $\surd$, but not mentioned & $\surd$, but not mentioned&$\surd$, but not mentioned &$\surd$, but not mentioned &$\surd$ \\
\hline
\centering
Multiple communication users & $\times$ & $\surd$ & $\surd$ &$\surd$ &$\surd$ & $\surd$\\
\hline
\centering
Unicast transmission & $\times$ & $\surd$ & $\surd$  &$\surd$ &$\surd$ & $\surd$\\
\hline
\centering
Multicast transmission & $\times$ & $\times$ & $\times$ &$\times$ &$\times$ &$\surd$ \\
\hline
\centering
The employment of NOMA & $\times$ & $\times$ & $\times$ &$\times$ &$\times$ &$\surd$ \\
\Xhline{1.4pt}
\end{tabular}
}
\end{center}
\label{table:structure2}
\end{table*}
\indent Against the above background, we propose the novel concept of NOMA-aided joint radar and multicast-unicast communication (Rad-MU-Com) and investigate the corresponding radar and communication BF design problems. The main contributions of this paper can be summarized below, which are boldly and explicitly contrasted to the relevant state-of-the-art in Table \ref{table:structure2}.
\begin{itemize}
  \item We investigate a NOMA-aided joint Rad-MU-Com system, which consists of two types of users, namely the radar-centric users (R-user) and the communication-centric users (C-user). By employing power-domain NOMA, a \emph{double spectrum sharing} operation is facilitated, where the MIMO DFRC base station (BS) transmits the mixed multicast and unicast messages to the R- and C-users, while detecting the R-user target using the transmitted superimposed communication signals.
  \item For the system supporting a single pair of R- and C-users, we first propose a beamformer-based NOMA (BB NOMA)-aided joint Rad-MU-Com framework, where the multicast and unicast messages are transmitted via different BFs. Based on this, we formulate a BF design problem for the maximization of the unicast rate, subject to both the multicast rate requirement and to the radar beam pattern accuracy achieved. To solve the resultant non-convex problem, we develop an efficient penalty-based iterative algorithm for finding a stationary point of the original optimization problem.
  \item For the system supporting multiple pairs of R- and C-users, we further propose a cluster-based NOMA (CB NOMA)-aided joint Rad-MU-Com framework, where the multicast and unicast messages for a R\&C pair are superimposed at different power levels and transmitted using a common BF. In this case, we formulate a joint BF design and power allocation problem for the maximization of the sum of the unicast rate, subject to specific constraints on the multicast rate of each R\&C pair and on the radar beam pattern accuracy achieved for the detection of multiple R-user targets. The joint optimization problem formulated is efficiently solved by conceiving another developed penalty-based iterative algorithm.
  \item Our numerical results show that the proposed NOMA-aided joint Rad-MU-Com schemes achieve a higher unicast performance than the benchmark schemes relying on conventional transmission strategies. Furthermore, the performance gain becomes more significant, when the constraint on the radar beam pattern is more relaxed. It also shows that the proposed NOMA-aided joint Rad-MU-Com schemes are capable of supporting multicast-unicast communication, while simultaneously constructing a high-quality radar beam pattern.
\end{itemize}
\subsection{Organization and Notation}
The rest of this paper is organized as follows. In Section II, a BB NOMA-aided joint Rad-MU-Com framework is conceived for a single pair of R- and C-users. Then, a unicast rate maximization problem is formulated, which is solved by our penalty-based algorithm. In Section III, a CB NOMA-aided joint Rad-MU-Com framework is designed for multiple pairs of R- and C-users, and a unicast sum rate maximization problem is formulated and solved by the penalty-based algorithm developed. Section IV provides numerical results for characterizing the proposed schemes compared to the relevant benchmark schemes. Finally, Section V concludes the paper.\\
\indent \emph{Notations:} Scalars, vectors, and matrices are denoted by lower-case, bold-face lower-case, and bold-face upper-case letters, respectively; ${\mathbb{C}^{N \times 1}}$ denotes the space of $N \times 1$ complex-valued vectors; ${{\mathbf{a}}^H}$ and $\left\| {\mathbf{a}} \right\|$ represent the conjugate transpose of vector ${\mathbf{a}}$; ${\mathcal{CN}}\left( {\mu,\sigma ^2} \right)$ denotes the distribution of a circularly symmetric complex Gaussian (CSCG) random variable with mean $\mu $ and variance ${\sigma ^2}$; ${{\mathbf{1}}}$ stands for the all-one vector; ${\rm {Rank}}\left( \mathbf{A} \right)$ and ${\rm {Tr}}\left( \mathbf{A} \right)$ denote the rank and the trace of matrix $\mathbf{A}$, respectively; ${\rm {Diag}}\left( \mathbf{A} \right)$ represents a vector whose elements are extracted from the main diagonal elements of matrix $\mathbf{A}$; ${{\mathbf{A}}} \succeq 0$ indicates that $\mathbf{A}$ is a positive semidefinite matrix; ${\mathbb{H}^{N}}$ denotes the set of all $N$-dimensional complex Hermitian matrices. ${\left\| {\mathbf{A}} \right\|_*}$, ${\left\| {\mathbf{A}} \right\|_2}$, and ${\left\| {\mathbf{A}} \right\|_F}$ are the nuclear norm, spectral norm, and Frobenius norm of matrix $\mathbf{A}$, respectively.
\section{BB NOMA-Aided Joint Rad-MU-Com System}
\subsection{System Model}
\begin{figure}[!ht]
  \centering
  \includegraphics[width=3in]{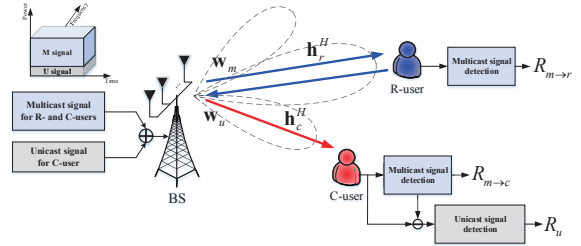}\\
  \caption{Illustration of the BB NOMA-aided joint Rad-MU-Com system.}\label{beamformer}
\end{figure}
As illustrated in Fig. \ref{beamformer}, a MIMO DFRC system is considered, which consists of a single $N$-antenna DFRC BS, a single-antenna R-user, and a single-antenna C-user. In contract to existing work~\cite{protocol,low,radio,security,Pareto,Optimal_Jie}, where the DFRC BS detects the R-user located within the angles of interest, while only communicating with the C-user, we consider mixed multicast-unicast transmission. To be more specific, two different types of messages have to be sent by the BS, one for both the R- and C-users, namely the multicast signal, while the unicast signal is only intended for the C-user. It is worth mentioning that this mixed multicast-unicast transmission represents the evolution of DFRC from \emph{isolation} to \emph{integration}. For instance, the multicast signal can be employed for broadcasting group-oriented system configurations and automatic software updates, which are requested both by the R- and C-users. By contrast, the unicast signal consists of personalized voice and video traffic intended for the C-user, which is not relevant for the R-user. To support this novel concept for DFRC, we propose a BB NOMA-aided joint Rad-MU-Com framework. In the following, the communication model and radar model of the proposed system will be introduced.
\subsubsection{BB NOMA-aided MU-Communication Model} For supporting our mixed multicast-unicast based MIMO DFRC system, the BB NOMA scheme of~\cite{multiple} is employed. Explicitly, the BS employs different BFs for transmitting the multicast signal intended for both R- and C-users and for the unicast signal only intended for the C-user, where the pair of signals are multiplexed in the power domain. Let $s_{m}\left[ n \right]$ and $s_{u}\left[ n \right]$ denote the multicast signal and the unicast signal at the time index $n$, respectively. Therefore, the corresponding transmitted superimposed signal at the $n$th time index is given by
\begin{align}\label{transmit signal}
{\mathbf{x}}_1\left[ n \right] = {{\mathbf{w}}_{m}}s_{m}\left[ n \right] + {{\mathbf{w}}_u}s_{u}\left[ n \right],
\end{align}
where ${{\mathbf{w}}_{m}} \in {{\mathbb{C}}^{N \times 1}}$ and ${{\mathbf{w}}_u}\in {{\mathbb{C}}^{N \times 1}}$ represents the BFs designed for transmitting the multicast and unicast information symbols, respectively. Without loss of generality, we assume that the multicast and unicast signals are statistically independent of each other and we have ${s_{m}}\left[ n \right] \sim {\mathcal{C}\mathcal{N}}\left( {0,1} \right)$ and ${s_u}\left[ n \right] \sim {\mathcal{C}\mathcal{N}}\left( {0,1} \right)$. Let ${{\mathbf{h}_r^H}}\in {{\mathbb{C}}^{1 \times N}}$ and ${{\mathbf{h}_c^H}}\in {{\mathbb{C}}^{1 \times N}}$ denote the BS-R-user channel and the BS-C-user channel, respectively. In this paper, we assume that the CSI can be perfectly estimated to study the maximum performance gain of the proposed NOMA-aided joint Rad-MU-Com system. The robust design with imperfect CSI is beyond the scope of this work and it is left for our future work. For the R- and C-users, the signal received at time index $n$ can be respectively expressed as follows:
\begin{align}\label{received signal T}
y_r\left[ n \right] = {\mathbf{h}_r^H}\left( {{{\mathbf{w}}_{m}}s_{m}\left[ n \right] + {{\mathbf{w}}_{u}}s_{u}\left[ n \right]} \right) + z_r\left[ n \right],
\end{align}
\begin{align}\label{received signal C}
y_c\left[ n \right] = {\mathbf{h}_c^H}\left( {{{\mathbf{w}}_{m}}s_{m}\left[ n \right] + {{\mathbf{w}}_{u}}s_{u}\left[ n \right]} \right) + z_c\left[ n \right],
\end{align}
where $z_r\left[ n \right]\sim {\mathcal{CN}}\left( {0,\sigma_r ^2} \right)$ and $z_c\left[ n \right]\sim {\mathcal{CN}}\left( {0,\sigma_c ^2} \right)$ denote the additive white Gaussian noise (AWGN) of the R- and C-users at the time index $n$, respectively. Similar to the ``strong'' user of conventional twin-user downlink NOMA transmission \cite{Liu2017}, the multicast signal is detected first at the C-user. Then, the remodulated multicast signal is subtracted from the composite received signal, automatically leaving the interference-free decontaminated unicast signal behind. As a result, the achievable rate for the multicast message at the C-user is given by
\begin{align}\label{CT rate C}
{R_{m \to c}} = {\log _2}\left( {1 + \frac{{{{\left| {{\mathbf{h}}_c^H{{\mathbf{w}}_{m}}} \right|}^2}}}{{{{\left| {{\mathbf{h}}_c^H{{\mathbf{w}}_u}} \right|}^2} + \sigma _c^2}}} \right).
\end{align}
After subtracting the remodulated multicast signal from the composite received signal by SIC, the achievable rate for the unicast signal at the C-user is given by
\begin{align}\label{C rate C}
{R_{u}} = {\log _2}\left( {1 + \frac{{{{\left| {{\mathbf{h}}_c^H{{\mathbf{w}}_u}} \right|}^2}}}{{\sigma _c^2}}} \right).
\end{align}
Similar to the ``weak'' user in the conventional twin-user downlink NOMA transmission~\cite{Liu2017}, the R-user directly detects the multicast signal by treating the unicast signal as noise. Therefore, the achievable rate for the multicast message can be expressed as
\begin{align}\label{CT rate T}
{R_{m \to r}} = {\log _2}\left( {1 + \frac{{{{\left| {{\mathbf{h}}_r^H{{\mathbf{w}}_{m}}} \right|}^2}}}{{{{\left| {{\mathbf{h}}_r^H{{\mathbf{w}}_u}} \right|}^2} + \sigma _r^2}}} \right).
\end{align}
The rate of the multicast signal is limited by the lower one of the pair of communication rates, which is given by
\begin{align}\label{CT rate}
{R_{m}} = \min \left\{ {{R_{m \to c}},{R_{m \to r}}} \right\}.
\end{align}
\subsubsection{BB NOMA-aided Radar Detection Model} According to \cite{protocol}, the above superimposed communication signals can also be exploited as radar probing waveforms, i.e., each transmitted information symbol can also be considered as a snapshot of a radar pulse. Therefore, the radar beam pattern design is equivalent to the design of the covariance matrix of the transmitted signal, ${\mathbf{x}}_1\left[ n \right]$, which is given by
\begin{align}\label{covariance matrix}
{\mathbf{R}_1} = {\mathbb{E}}\left[ {{\mathbf{x}}_1\left[ n \right]{{\mathbf{x}}_1^H}\left[ n \right]} \right] = {{\mathbf{w}}_{m}}{\mathbf{w}}_{m}^H + {{\mathbf{w}}_u}{\mathbf{w}}_u^H.
\end{align}
Then, the transmit beam pattern used for radar detection can be expressed as
\begin{align}\label{beam pattern}
P\left( \theta  \right) = {{\bm{\alpha}} ^H}\left( \theta  \right){\mathbf{R}_1}{\bm{\alpha}} \left( \theta  \right),
\end{align}
where ${\bm{\alpha}} \left( \theta  \right) = {\left[ {1,{e^{j\frac{{2\pi d}}{\lambda }\sin \theta }}, \ldots ,{e^{j\frac{{2\pi d}}{\lambda }\left( {N - 1} \right)\sin \theta }}} \right]^T}\in {{\mathbb{C}}^{N \times 1}}$ denotes the steering vector of the transmit antenna array, $\theta $ is the detection angle, $d$ represents the antenna spacing, and $\lambda$ is the carrier wavelength.
\begin{remark}\label{benefits}
\emph{The main benefits of the proposed BB NOMA-aided joint Rad-MU-Com framework can be summarized as follows. Firstly, the employment of NOMA ensures the quality of the unicast transmission (which is usually data-hungry) for the C-user, since the inter-signal interference is canceled by SIC\footnote{The employment of SIC introduces additional signal processing complexity. This, however, enables the proposed scheme to achieve a significant performance gain, see Section IV for details.}, see \eqref{C rate C}. Secondly, despite the presence of interference, the rate-requirement of both the R- and C-users can be readily guaranteed as a benefit of the power sharing provided by NOMA, see \eqref{CT rate C} and \eqref{CT rate T}. Thirdly, the different BFs used in our BB NOMA structure provide additional degrees-of-freedom (DoFs) for our radar beam pattern design, see \eqref{covariance matrix}. Last but not least, NOMA facilitates \emph{double spectrum sharing} between both the multicast and unicast as well as between radar and communication systems, thus further enhancing the SE.}
\end{remark}
\begin{remark}\label{bechmark schems}
\emph{The joint Rad-MU-Com concept may also be facilitated by existing conventional transmission schemes. For example, the multicast and unicast signals can be successively transmitted via different time slots while detecting the R-user target, namely by a time division multiple access (TDMA) based Rad-MU-Com system. Moreover, the multicast and unicast information can be transmitted via conventional BFs dispensing with SIC~\cite{protocol,low,radio,security,Pareto,Optimal_Jie}, while detecting the R-user target, namely by a CBF-No-SIC based Rad-MU-Com system. These options will serve as the benchmark schemes in our performance comparisons of Section IV.}
\end{remark}
\subsection{Problem Formulation}
Before formulating the associated optimization problem, we first introduce the concept of the ideal radar beam pattern, which can be obtained by solving the following least-squares problem~\cite{protocol,idea_pattern}:
\begin{subequations}\label{ideal beam pattern design}
\begin{align}
\mathop {\min }\limits_{\delta ,{{\mathbf{R}}_0}} &\;\Delta \left( {{{\mathbf{R}}_0},\delta } \right) \triangleq {\sum\nolimits_{m = 1}^M {\left| {\delta {P^*}\left( {{\theta _m}} \right) - {{\bm{\alpha}} ^H}\left( {{\theta _m}} \right){{\mathbf{R}}_0}{\bm{\alpha}} \left( {{\theta _m}} \right)} \right|} ^2}  \\
\label{average power}{\rm{s.t.}}\;\;&{\rm{Tr}}\left( {{{\mathbf{R}}_0}} \right) = {P_{\max }},\\
\label{SEMI}&{\mathbf{R}}_0 \succeq 0,{\mathbf{R}}_0 \in {{\mathbb{H}}^N},\\
\label{scale}&\delta \ge 0,
\end{align}
\end{subequations}
where $\left\{ {{\theta _m}} \right\}_{m = 1}^M$ denotes an angular grid covering the detector's angular range in $\left[ { - \frac{\pi }{2},\frac{\pi }{2}} \right]$, ${\bm{\alpha}} \left( {{\theta _m}}\right)$ is the corresponding steering vector, ${P^*}\left( {{\theta _m}}\right)$ represents the desired ideal beam pattern gain at ${{\theta _m}}$, $\delta$ is a scaling factor, $P_{\max}$ is the maximum transmit power budget at the MIMO DFRC BS\footnote{In this paper, the total power constraint is considered for the MIMO DFRC BS~\cite{security,Optimal_Jie}, which provides high DoFs for the BF design than the per-antenna power constraint.}, and ${\mathbf{R}_0}$ is the waveform's covariance matrix, when only the MIMO radar is considered. It can be readily verified that the ideal radar beam pattern design problem of \eqref{ideal beam pattern design} is convex, which can be efficiently solved. Let ${\mathbf{R}_0^*}$ and $\delta^*$ denote the optimal solutions of \eqref{ideal beam pattern design}. The corresponding objective function value $\Delta \left( {{\mathbf{R}}_0^*,{\delta ^*}} \right)$ characterizes the minimum beam pattern error between the desired ideal beam pattern gain and the radar-only beam pattern gain. However, for supporting both the communication and radar functions in the MIMO DFRC system considered, a radar performance loss will occur. In the following, $\Delta \left( {{\mathbf{R}}_0^*,{\delta ^*}} \right)$ will be used as a performance benchmark for quantifying the radar performance loss in the joint Rad-MU-Com system design.\\
\indent Given our BB NOMA-aided joint Rad-MU-Com framework and the radar performance benchmark $\Delta \left( {{\mathbf{R}}_0^*,{\delta ^*}} \right)$, we aim for maximizing the unicast rate achieved at the C-user, while satisfying the minimum rate requirement of multicast communication at both the R- and C-users as well as achieving the desired beam pattern for radar detection. The resultant optimization problem can be formulated as follows:
\begin{subequations}\label{beamformer based 0}
\begin{align}
&\mathop {\max }\limits_{{{\mathbf{w}}_{m}},{{\mathbf{w}}_u},{{\mathbf{R}}_1}} \;{R_{u}} \\
\label{CT QoS 2}{\rm{s.t.}}\;\;&{R_{m}} \ge {{\overline R}_{m}},\\
\label{radar QoS}&\frac{{\Delta \left( {{{\mathbf{R}}_1},{\delta ^*}} \right) - \Delta \left( {{\mathbf{R}}_0^*,{\delta ^*}} \right)}}{{\Delta \left( {{\mathbf{R}}_0^*,{\delta ^*}} \right)}} \le {{\overline \gamma }_b},\\
\label{power Communication}&{\rm{Tr}}\left( {{{\mathbf{R}}_1}} \right) = {P_{\max }},
\end{align}
\end{subequations}
where $\overline R_{m}$ represents the minimum rate requirement of multicast, and $\overline \gamma_b$ is the maximum tolerable radar beam pattern mismatch ratio between the beam pattern error achieved in the joint Rad-MU-Com system (i.e., $\Delta \left( {{\mathbf{R}}_1,{\delta ^*}} \right)$) and the minimum one (i.e., $\Delta \left( {{\mathbf{R}}_0^*,{\delta ^*}} \right)$) obtained by the radar-only system.
\subsection{Proposed Solution}
\begin{figure}[!t]
  \centering
  \includegraphics[width=2.2in]{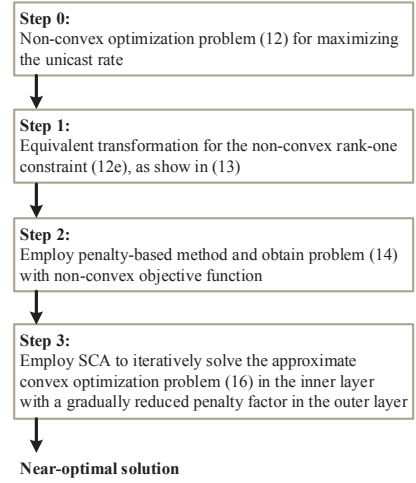}\\
  \caption{Illustration of the key steps in solving problem \eqref{beamformer based 1}.}\label{BB}
\end{figure}

The main challenge in solving problem \eqref{beamformer based 0} is that the objective function and the left-hand-side (LHS) is not concave with respect to the optimization variables. To address this issue, we define ${{\mathbf{W}}_{m}} = {{\mathbf{w}}_{m}}{\mathbf{w}}_{m}^H$ and ${{\mathbf{W}}_{u}} = {{\mathbf{w}}_{u}}{\mathbf{w}}_{u}^H$, which satisfy that ${{\mathbf{W}}_{m}}\succeq 0$, ${{\mathbf{W}}_{u}}\succeq 0$, ${\rm{Rank}}\left( {{{\mathbf{W}}_{m}}} \right)=1$, and ${\rm{Rank}}\left( {{{\mathbf{W}}_u}} \right) = 1$. Then, problem \eqref{beamformer based 0} can be reformulated as follows:
\begin{subequations}\label{beamformer based 1}
\begin{align}
&\mathop {\max }\limits_{{{\mathbf{W}}_{m}},{{\mathbf{W}}_u},{\mathbf{R}_1}} \;{\log _2}\left( {1 + \frac{{{\rm{Tr}}\left( {{{\mathbf{H}}_c}{{\mathbf{W}}_u}} \right)}}{{\sigma _c^2}}} \right)\\
\label{CT C QoS}{\rm{s.t.}}\;\;&{\rm{Tr}}\left( {{{\mathbf{H}}_c}{{\mathbf{W}}_{m}}} \right) - {\overline \gamma  _{m}}{\rm{Tr}}\left( {{{\mathbf{H}}_c}{{\mathbf{W}}_u}} \right) - {\overline \gamma  _{m}}\sigma _c^2 \ge 0,\\
\label{CT T QoS}&{\rm{Tr}}\left( {{{\mathbf{H}}_r}{{\mathbf{W}}_{m}}} \right) - {\overline \gamma  _{m}}{\rm{Tr}}\left( {{{\mathbf{H}}_r}{{\mathbf{W}}_u}} \right) - {\overline \gamma  _{m}}\sigma _r^2 \ge 0,\\
\label{rank 1}&{{\mathbf{W}}_{m}}, {{\mathbf{W}}_{u}}\succeq 0, {{\mathbf{W}}_{m}}, {{\mathbf{W}}_{u}}\in {{\mathbb{H}}^N},\\
\label{rank}&{\rm{Rank}}\left( {{{\mathbf{W}}_{m}}} \right)=1, {\rm{Rank}}\left( {{{\mathbf{W}}_u}} \right) = 1,\\
\label{constraints beamformer 1}&\eqref{radar QoS},\eqref{power Communication},
\end{align}
\end{subequations}
where \eqref{CT C QoS} and \eqref{CT T QoS} are arranged from \eqref{CT QoS 2}. Furthermore, we defined ${{\mathbf{H}}_c} \triangleq {{\mathbf{h}}_c}{\mathbf{h}}_c^H$, ${{\mathbf{H}}_r} \triangleq {{\mathbf{h}}_r}{\mathbf{h}}_r^H$,  and ${\overline \gamma  _{m}} = {2^{{{\overline R}_{m}}}} - 1$. Now, the non-convexity of the reformulated problem \eqref{beamformer based 1} only lies in the rank-one constraint \eqref{rank}. To tackle this obstacle, a popular technique is to use semidefinite relaxation (SDR)~\cite{protocol}. Explicitly, we firstly solve the problem by ignoring the rank-one constraint and then apply the Gaussian randomization method for constructing a rank-one solution, if the resultant solution is not of rank-one. The advantage of employing the SDR is that the computational complexity may be low, since the relaxed problem only has to be solved once. However, considerable performance erosion may occur due to the reconstruction. On the other hand, it cannot be guaranteed that the reconstructed rank-one solution is still feasible in terms of satisfying all other constraints of the original problem (e.g., \eqref{radar QoS}, \eqref{CT C QoS}, and \eqref{CT T QoS}). As a remedy, a double-layer penalty-based iterative algorithm is proposed for gradually finding a near-optimal rank-one solution. Before introducing the detailed manipulations, the key steps of the proposed solution for solving problem \eqref{beamformer based 1} are depicted in Fig. \ref{BB}. \\
\indent To begin with, the non-convex rank-one constraint \eqref{rank} is equivalent to the following equality constraints:
\begin{subequations}\label{DC rank}
\begin{align}\label{DC rank 1}&{\left\| {{{\mathbf{W}}_{m}}} \right\|_*} - {\left\| {{{\mathbf{W}}_{m}}} \right\|_2} = 0,\\
\label{DC rank 2}&{\left\| {{{\mathbf{W}}_{u}}} \right\|_*} - {\left\| {{{\mathbf{W}}_{u}}} \right\|_2} = 0,
\end{align}
\end{subequations}
where ${\left\| \cdot  \right\|_ * }$ and ${\left\| \cdot \right\|_2} $ denote the nuclear norm and spectral norm of the matrix, respectively. Let us consider ${\mathbf{W}}_{m}$ as an example. It can be verified that, for any ${{\mathbf{W}}_{m}} \in {{\mathbb{H}}^N}$ and ${{\mathbf{W}}_{m}} \succeq 0$, the above equality constraint is only satisfied, when the matrix ${\mathbf{W}}_{m}$ is of rank-one. Otherwise, we always have ${\left\| {{{\mathbf{W}}_{m}}} \right\|_*} - {\left\| {{{\mathbf{W}}_{m}}} \right\|_2} > 0$.\\
\indent To solve problem \eqref{beamformer based 1}, we employ the penalty-based method of~\cite{penalty} by introducing the transformed equality constraints for ${\mathbf{W}}_{m}$ and ${\mathbf{W}}_{u}$ as a penalty term into the objective function of \eqref{beamformer based 1}, yielding the following optimization problem:
\begin{subequations}\label{beamformer based 2}
\begin{align}
\mathop {\min }\limits_{{{\mathbf{W}}_m},{{\mathbf{W}}_u},{{\mathbf{R}}_1}}&  \!- \!{\rm{Tr}}\left( {{{\mathbf{H}}_c}{{\mathbf{W}}_u}} \right) \!+\! \frac{1}{{{\eta _1}}}\sum\nolimits_{i \in \left\{ {m,u} \right\}}\! {\left( {{{\left\| {{{\mathbf{W}}_i}} \right\|}_*}\! -\! {{\left\| {{{\mathbf{W}}_i}} \right\|}_2}} \right)} \\
\label{constraints beamformer 2}{\rm{s.t.}}\;\;&\eqref{radar QoS},\eqref{power Communication},\eqref{CT C QoS}-\eqref{rank 1},
\end{align}
\end{subequations}
where $\eta_1  > 0$ is the penalty factor, which penalizes the violation of the equality constraints \eqref{DC rank 1} and \eqref{DC rank 2}, i.e., when ${\mathbf{W}}_{m}$ and ${\mathbf{W}}_{u}$ are not of rank-one. Since the maximization of ${R_{u}}$ is equivalent to maximizing the corresponding received signal strength of ${\rm{Tr}}\left( {{{\mathbf{H}}_c}{{\mathbf{W}}_u}} \right)$, we drop the $\log$ function in the objective function of \eqref{beamformer based 2} for simplicity. Despite relaxing the equality constraints in problem \eqref{beamformer based 2}, it may be readily verified that the solutions obtained will always satisfy the equality constraints (i.e., have rank-one matrices), when $\frac{1}{\eta_1 } \to  + \infty $ ($\eta_1  \to 0$). This is because if the rank of any of the obtained matrix solutions $\left\{ {{{\mathbf{W}}_m},{{\mathbf{W}}_u}} \right\}$ at $\frac{1}{\eta_1 } \to  + \infty $ is larger than one, the corresponding objective function value will be infinitely large. In this case, we can have rank-one matrix solutions satisfying the equality constraints \eqref{DC rank} to render the penalty term zero, which in turn achieves a finite objective function value. Therefore, problems \eqref{beamformer based 1} and \eqref{beamformer based 2} are equivalent when $\frac{1}{\eta_1 } \to  + \infty $. However, if we firstly initialize $\eta_1 $ with a sufficiently small value, the objective function's value of \eqref{beamformer based 2} tends to be dominated by the penalty term introduced, thus significantly degrading the efficiency of maximizing ${\rm{Tr}}\left( {{{\mathbf{H}}_c}{{\mathbf{W}}_u}} \right)$. To facilitate efficient optimization, we can initialize $\eta_1 $ with a sufficiently large value to find a good starting point, and then gradually reduce $\eta_1 $ to a sufficiently small value. As a result, feasible rank-one matrix solutions associated with a near-optimal performance can eventually be obtained. In the following, we will present the details of the double-layer penalty-based algorithm for solving problem \eqref{beamformer based 2}. In the inner layer, the optimization problem for a given $\eta_1 $ is solved iteratively by employing successive convex approximation (SCA)~\cite{SCA} until convergence is reached. In the outer layer, the penalty factor, $\eta_1 $, is gradually reduced from a sufficiently large value to a sufficiently small one.
\subsubsection{Inner Layer: Solving Problem \eqref{beamformer based 2} for A Given $\eta_1 $} Note that for a given $\eta_1 $, the non-convexity of \eqref{beamformer based 2} manifests itself in that the second term of each penalty term is non-convex, i.e., $ - {\left\| {{{\mathbf{W}}_{m}}} \right\|_2}$ and $ - {\left\| {{{\mathbf{W}}_u}} \right\|_2}$. However, they are concave functions with respect to both ${{{\mathbf{W}}_{m}}}$ and ${{{\mathbf{W}}_{u}}}$. By exploiting the first-order Taylor expansion, their upper bounds can be respectively expressed as follows:
\begin{subequations}
\begin{align}\label{W1 uppder bound}&
\begin{gathered}
   - {\left\| {{{\mathbf{W}}_m}} \right\|_2} \le \overline {\mathbf{W}} _m^n \triangleq  - {\left\| {{\mathbf{W}}_m^n} \right\|_2} \hfill \\
   - {\rm{Tr}}\left[ {{{\mathbf{v}}_{\max }}\left( {{\mathbf{W}}_m^n} \right){{\mathbf{v}}_{{{\max }}}^H}\left( {{\mathbf{W}}_m^n} \right)\left( {{{\mathbf{W}}_m} - {\mathbf{W}}_m^n} \right)} \right], \hfill \\
\end{gathered} \\
\label{W2 uppder bound}&
\begin{gathered}
   - {\left\| {{{\mathbf{W}}_u}} \right\|_2} \le \overline {\mathbf{W}} _u^n \triangleq  - {\left\| {{\mathbf{W}}_u^n} \right\|_2} \hfill \\
   - {\rm{Tr}}\left[ {{{\mathbf{v}}_{\max }}\left( {{\mathbf{W}}_u^n} \right){{\mathbf{v}}_{{{\max }}}^H}\left( {{\mathbf{W}}_u^n} \right)\left( {{{\mathbf{W}}_u} - {\mathbf{W}}_u^n} \right)} \right], \hfill \\
\end{gathered}
\end{align}
\end{subequations}
where ${{\mathbf{W}}_{m}^n}$ and ${{\mathbf{W}}_{u}^n}$ denote given points during the $n$th iteration of the SCA method, while ${{{\mathbf{v}}_{\max }}\left( {{\mathbf{W}}_{m}^n} \right)}$ and ${{{\mathbf{v}}_{\max }}\left( {{\mathbf{W}}_{u}^n} \right)}$ represent the eigenvector corresponding to the largest eigenvalue of ${{\mathbf{W}}_{m}^n}$ and ${{\mathbf{W}}_{u}^n}$, respectively.\\
\indent Accordingly, by exploiting the upper bounds obtained, problem \eqref{beamformer based 2} can be approximated by the following convex optimization problem:
\begin{subequations}\label{beamformer based 3}
\begin{align}
\mathop {\min }\limits_{{{\mathbf{W}}_m},{{\mathbf{W}}_u},{{\mathbf{R}}_1}}& \! -\! {\rm{Tr}}\left( {{{\mathbf{H}}_c}{{\mathbf{W}}_u}} \right) \!+\! \frac{1}{{{\eta _1}}}\sum\nolimits_{i \in \left\{ {m,u} \right\}} \!{\left( {{{\left\| {{{\mathbf{W}}_i}} \right\|}_*} - \overline {\mathbf{W}} _i^n} \right)} \\
\label{constraints beamformer 3}{\rm{s.t.}}\;\;&\eqref{radar QoS},\eqref{power Communication},\eqref{CT C QoS}-\eqref{rank 1}.
\end{align}
\end{subequations}
The above convex optimization problem can be efficiently solved by using existing standard convex problem solvers such as CVX~\cite{cvx}. Therefore, for a given $\eta_1$, problem \eqref{beamformer based 3} is iteratively solved until the fractional reduction of the objective function's value in \eqref{beamformer based 3} falls below the predefined threshold, $\epsilon_i$, when convergence is declared.
\subsubsection{Outer Layer: Reducing the Penalty Factor $\eta_1 $}
In order to satisfy the equality constraints \eqref{DC rank 1} and \eqref{DC rank 2}, in the outer layer, we gradually update the value of $\eta_1$ towards a sufficiently small value as follows:
\begin{align}\label{Update}
\eta_1  = \varepsilon \eta_1 ,0 < \varepsilon  < 1,
\end{align}
where $\varepsilon$ is a constant scaling factor, which has to be carefully selected for striking performance vs. complexity trade-off. For example, a larger $\varepsilon$ allows us to explore more potential candidate solutions, thus ultimately achieving a higher final performance. This, however, in turn requires more outer iterations hence imposing a higher complexity.
\subsubsection{Overall Algorithm and Complexity Analysis} Based on the above discussion, the proposed double-layer penalty-based procedure is summarized in \textbf{Algorithm 1}. The termination of the proposed algorithm depends on the violation of the equality constraints, which is expressed as follows:
\begin{align}\label{termination}
\max \left\{ {{{\left\| {{{\mathbf{W}}_{m}}} \right\|}_*} - {{\left\| {{{\mathbf{W}}_{m}}} \right\|}_2},{{\left\| {{{\mathbf{W}}_u}} \right\|}_*} - {{\left\| {{{\mathbf{W}}_u}} \right\|}_2}} \right\} \le \epsilon_o ,
\end{align}
where $\epsilon_o$ represents the maximum tolerable value. Upon reducing $\eta_1$, the equality constraints will finally be satisfied at an accuracy of $\epsilon_o$. For the given $\eta_1 $ in the inner layer, the objective function value of \eqref{beamformer based 3} is monotonically non-increasing over each iteration and the unicast rate is upper-bounded due to the limited transmit power of the BS. Therefore, the proposed double-layer penalty-based algorithm is guaranteed to converge to a stationary point of the original problem \eqref{beamformer based 0}~\cite{SCA}.\\
\indent The main complexity of \textbf{Algorithm 1} arises from iteratively solving problem \eqref{beamformer based 3}. Since problem \eqref{beamformer based 3} is a standard semidefinite program (SDP), the corresponding complexity is of the order of ${\mathcal{O}}\left( {2{N^{3.5}}} \right)$~\cite{Luo}. Therefore, the overall complexity of \textbf{Algorithm 1} is ${{\mathcal{O}}}\left( {{I_o^1}{I_i^1}\left( {2{N^{3.5}}} \right)} \right)$, where ${I_{i}^1}$ and ${I_{o}^1}$ denote the number of inner and outer iterations required for the convergence of \textbf{Algorithm 1}, respectively.
\begin{algorithm}[!t]\label{method1}
\caption{Proposed double-layer penalty-based algorithm for solving problem \eqref{beamformer based 0}}
\begin{algorithmic}[1]
\STATE {Initialize feasible points ${{\mathbf{W}}_{m}^0}$ and ${{\mathbf{W}}_{u}^0}$ as well as the penalty factor $\eta_1$.}
\STATE {\bf repeat: outer layer}
\STATE \quad Set iteration index $n=0$ for inner layer.
\STATE \quad {\bf repeat: inner layer}
\STATE \quad\quad For given ${{\mathbf{W}}_{m}^n}$ and ${{\mathbf{W}}_{u}^n}$, solve the convex problem \eqref{beamformer based 3} and the solutions obtained are denoted by ${{\mathbf{W}}_{m}^*}$ and ${{\mathbf{W}}_{m}^*}$.
\STATE \quad\quad ${\mathbf{W}}_{m}^{n + 1} = {\mathbf{W}}_{m}^*$, ${\mathbf{W}}_{u}^{n + 1} = {\mathbf{W}}_{u}^*$, and $n=n+1$.
\STATE \quad {\bf until} the fractional reduction of the objective function value falls below a predefined threshold $\epsilon_i >0$.
\STATE \quad ${\mathbf{W}}_{m}^0 = {\mathbf{W}}_{m}^*$, ${\mathbf{W}}_{u}^0 = {\mathbf{W}}_{u}^*$.
\STATE \quad Update $\eta_1  = \varepsilon \eta_1$.
\STATE {\bf until} the constraint violation falls below a maximum tolerable threshold $\epsilon_o >0$.
\end{algorithmic}
\end{algorithm}
\section{CB NOMA-Aided Joint Rad-MU-Com System}
The joint Rad-MU-Com system of Section II serves a single pair of R- and C-users (referred to as a R\&C pair), both of which require the same multicast signals\footnote{In this paper, we assume that the number of R- and C-users is the same and each R\&C pair consists of a pair of R- and C-users. The problem for the system having different numbers of R- and C-users leaves as our future work.}. However, in practice, there may be multiple R\&C pairs in the joint Rad-MU-Com system. In this case, the DFRC BS has to transmit multiple mixed multicast and unicast messages, while detecting multiple R-user targets. Hence, in this section we propose a CB NOMA-aided joint Rad-MU-Com framework. The key idea is to employ a common BF for conveying both the multicast and unicast messages via NOMA to the R- and C-users in one pair. Then multiple BFs are employed simultaneously for jointly detecting multiple Ruser targets in the system.
\subsection{System Model}
\begin{figure}[!ht]
  \centering
  \includegraphics[width=3in]{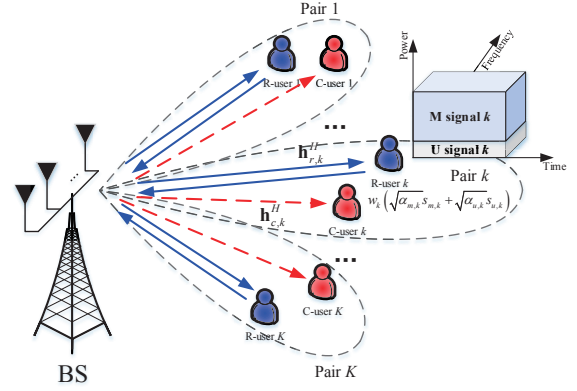}\\
  \caption{Illustration of the CB NOMA-aided joint Rad-MU-Com system.}\label{cluster}
\end{figure}
As shown in Fig. \ref{cluster}, we now consider a joint Rad-MU-Com system having $K>1$ R\&C pairs. For each pair, the DFRC BS carries out the mixed multicast and unicast transmission, i.e., transmitting a multicast message to both the R- and C-users within the same pair, while only delivering the unicast message to the C-user. Meanwhile, the DFRC BS also has to detect the $K$ R-user targets located at different angles of interest.
\subsubsection{CB NOMA-aided MU-Communication Model} In contrast to the BB NOMA structure of Section II, where the multicast and unicast messages are respectively transmitted via different BFs, the twinned messages are multiplexed in the power domain and they are delivered via a common BF for the intended pair~\cite{multiple}. Let $s_{m,k}\left[ n \right]$ and $s_{u,k}\left[ n \right]$ denote the multicast signal and the unicast signal intended for the $k$th pair at the time index $n$, respectively, where $k \in {\mathcal{K}} \triangleq \left\{ {1,2, \ldots ,K} \right\}$. The BF constructed for the $k$th pair is denoted by ${{\mathbf{w}}_k}\in {{\mathbb{C}}^{N \times 1}}$. Therefore, the signal transmitted to the $K$ pairs at the $n$th time index is given by
\begin{align}\label{transmit signal CB}
{{\mathbf{x}}_2}\left[ n \right] = \sum\nolimits_{k = 1}^K {{{\mathbf{w}}_k}\left( {\sqrt {\alpha _{m,k}} s_{m,k}\left[ n \right] + \sqrt {\alpha _{u,k}} s_{u,k}\left[ n \right]} \right)} ,
\end{align}
where $\alpha _{m,k} \ge 0$ and $\alpha _{u,k} \ge 0$ denote the power allocation factor of the multicast and unicast signals of the $k$th pair, respectively. Without loss of generality, we have $\alpha _{m,k} + \alpha _{u,k} = 1$. Let ${{\mathbf{h}_{r,k}^H}}\in {{\mathbb{C}}^{1 \times N}}$ and ${{\mathbf{h}_{c,k}^H}}\in {{\mathbb{C}}^{1 \times N}}$ denote the channels from the BS to the R- and C-users in the $k$th pair, respectively, which are assumed to be perfectly estimated. Accordingly, at the time index $n$, the signal received by the R- and C-users in the $k$th pair can be respectively expressed as follows
\begin{align}\label{received signal CB T}
\begin{gathered}
  {y_{r,k}}\left[ n \right] = \underbrace {{\mathbf{h}}_{r,k}^H{{\mathbf{w}}_k}\sqrt {{\alpha _{m,k}}} {s_{m,k}}\left[ n \right]}_{{\rm{desired\;multicast\;signal}}} + \underbrace {{\mathbf{h}}_{r,k}^H{{\mathbf{w}}_k}\sqrt {{\alpha _{u,k}}} {s_{u,k}}\left[ n \right]}_{{\rm{intra - pair\;interference}}} \hfill \\
   + \underbrace {{\mathbf{h}}_{r,k}^H\sum\nolimits_{i \ne k}^K {{{\mathbf{w}}_i}\left( {\sqrt {{\alpha _{m,i}}} {s_{m,i}}\left[ n \right] + \sqrt {{\alpha _{u,i}}} {s_{u,i}}\left[ n \right]} \right)}  + {z_{r,k}}\left[ n \right]}_{{\rm{inter - pair\;interference + noise}}}, \hfill \\
\end{gathered}
\end{align}
\begin{align}\label{received signal CB C}
\begin{gathered}
  {y_{c,k}}\left[ n \right] = \underbrace {{\mathbf{h}}_{c,k}^H{{\mathbf{w}}_k}\sqrt {{\alpha _{m,k}}} {s_{m,k}}\left[ n \right]}_{{\rm{desired\;multicast\;signal}}} + \underbrace {{\mathbf{h}}_{c,k}^H{{\mathbf{w}}_k}\sqrt {{\alpha _{u,k}}} {s_{u,k}}\left[ n \right]}_{{\rm{desired\;unicast\;signal}}} \hfill \\
   + \underbrace {{\mathbf{h}}_{c,k}^H\sum\nolimits_{i \ne k}^K {{{\mathbf{w}}_i}\left( {\sqrt {{\alpha _{m,i}}} {s_{m,i}}\left[ n \right] + \sqrt {{\alpha _{u,i}}} {s_{u,i}}\left[ n \right]} \right)}  + {z_{c,k}}\left[ n \right]}_{{\rm{inter - pair\;interference + noise}}}, \hfill \\
\end{gathered}
\end{align}
where $z_{r,k}\left[ n \right]\sim {\mathcal{CN}}\left( {0,\sigma_{r,k} ^2} \right)$ and $z_{c,k}\left[ n \right]\sim {\mathcal{CN}}\left( {0,\sigma_{c,k} ^2} \right)$ denote the AWGN of the R- and C-users in the $k$th pair at the time index $n$, respectively.\\
\indent Similarly, for each pair, downlink NOMA transmission is employed. The C-user firstly detects its intended multicast signal by treating the other signals as interference, and then detects its intended unicast signal after the SIC operation with the presence of the intra-pair interference. Therefore, the achievable rate for the intended multicast signal of the C-user in the $k$th pair is given by
\begin{align}\label{CT rate C k}
{R_{m \to c,k}}\! =\!  {\log _2}\! \left( \! {1\!  +\!  \frac{{{\alpha _{m,k}}{{\left| {{\mathbf{h}}_{c,k}^H{{\mathbf{w}}_k}} \right|}^2}}}{{{\alpha _{u,k}}{{\left| {{\mathbf{h}}_{c,k}^H{{\mathbf{w}}_k}} \right|}^2}\!\!   + \! \sum\nolimits_{i \ne k}^K {{{\left| {{\mathbf{h}}_{c,k}^H{{\mathbf{w}}_i}} \right|}^2}}\!\!    +\!  \sigma _{c,k}^2}}} \! \right).
\end{align}
After SIC, the achievable rate of the intended unicast signal at the $k$th C-user is given by
\begin{align}\label{C rate C k}
{R_{u,k}} = {\log _2}\left( {1 + \frac{{{\alpha _{u,k}}{{\left| {{\mathbf{h}}_{c,k}^H{{\mathbf{w}}_k}} \right|}^2}}}{{\sum\nolimits_{i \ne k}^K {{{\left| {{\mathbf{h}}_{c,k}^H{{\mathbf{w}}_i}} \right|}^2}}  + \sigma _{c,k}^2}}} \right).
\end{align}
Since the R-user of each pair only detects its intended multicast signal, its achievable rate in the $k$th pair is given by
\begin{align}\label{C rate T k}
{R_{m \to r,k}}\!  =\!  {\log _2}\! \left(\!  {1 \! +\!  \frac{{{\alpha _{m,k}}{{\left| {{\mathbf{h}}_{r,k}^H{{\mathbf{w}}_k}} \right|}^2}}}{{{\alpha _{u,k}}{{\left| {{\mathbf{h}}_{r,k}^H{{\mathbf{w}}_k}} \right|}^2} \! \! + \! \sum\nolimits_{i \ne k}^K {{{\left| {{\mathbf{h}}_{r,k}^H{{\mathbf{w}}_i}} \right|}^2}} \! \!  +\!  \sigma _{r,k}^2}}} \! \right).
\end{align}
Similarly, the overall multicast rate of the $k$th pair is given by
\begin{align}\label{CT rate k}
{R_{m,k}} = \min \left\{ {{R_{m \to c,k}},{R_{m \to r,k}}} \right\}.
\end{align}
\subsubsection{CB NOMA-aided Radar Detection Model} In this case, the covariance matrix of the transmitted signal, ${\mathbf{x}}_2\left[ n \right]$, is given by
\begin{align}\label{covariance matrix CB}
{\mathbf{R}_2} = {\mathbb{E}}\left[ {{\mathbf{x}}_2\left[ n \right]{{\mathbf{x}}_2^H}\left[ n \right]} \right] = \sum\nolimits_{k = 1}^K {{{\mathbf{w}}_k}{\mathbf{w}}_k^H}.
\end{align}
The transmit beam pattern constructed for radar detection can be obtained upon replacing ${\mathbf{R}_1}$ of \eqref{beam pattern} by ${\mathbf{R}_2}$.
\begin{remark}\label{cluster vs beamformer}
\emph{Note that the BB NOMA-aided joint Rad-MU-Com framework can also be employed for systems having $K>1$ R\&C pairs. To facilitate this design, each pair requires 2 BFs for respectively delivering the multicast and unicast messages, thus leading to a total of $2K$ BFs for our joint Rad-MU-Com system. Despite providing enhanced DoFs for system design, the resultant complexity may become excessive, when $K$ is large. By contrast, our CB NOMA-aided joint Rad-MU-Com framework only requires a total of $K$ BFs for achieving the same goal at a reduced complexity, which motivates us to exploit this design.}
\end{remark}
\subsection{Problem Formulation}
In this context, our aim is to maximize the sum of the unicast rate of all C-users, subject to the constraints on both the rate requirements of the multicast in each pair and on the mismatch between the achieved and the actual true beam pattern of the $K$ R-user targets. Therefore, the optimization problem can be formulated as follows:
\begin{subequations}\label{cluster based 0}
\begin{align}
\mathop {\max }\limits_{\left\{ {{{\mathbf{w}}_k},{\alpha _{m,k}},{\alpha _{u,k}}} \right\},{\mathbf{R}}_2} &\;\sum\nolimits_{k = 1}^K {{R_{u,k}}}\\
\label{cluster QoS 0}{\rm{s.t.}}\;\;&{R_{m,k}} \ge {{\overline R}_{m,k}},\forall k \in {\mathcal{K}},\\
\label{cluster radar QoS}&\frac{{\Delta \left( {{{\mathbf{R}}_2},{\delta ^*}} \right) - \Delta \left( {{\mathbf{R}}_0^*,{\delta ^*}} \right)}}{{\Delta \left( {{\mathbf{R}}_0^*,{\delta ^*}} \right)}} \le {{\overline \gamma }_b},\\
\label{cluster power Communication}&{\rm{Tr}}\left( {{{\mathbf{R}}_2}} \right) = {P_{\max }},\\
\label{power allocation 1}&{\alpha _{m,k}} + {\alpha _{u,k}} = 1,\forall k \in {\mathcal{K}},\\
\label{power allocation 2}&{\alpha _{m,k}} \ge 0,{\alpha _{u,k}} \ge 0,\forall k \in {\mathcal{K}},
\end{align}
\end{subequations}
where ${{\overline R}_{m,k}}$ denotes the minimum required multicast rate of the $k$th pair and ${{\mathbf{R}}_0^*}$ represents the actual true radar beam pattern, which can be obtained by solving problem \eqref{ideal beam pattern design} for detecting $K$ R-users. Problem \eqref{cluster based 0} is a non-convex optimization problem due to the non-convex objective function and the non-convex multicast rate constraint \eqref{cluster QoS 0}, where the power allocation factors, $\left\{ {{\alpha _{m,k}},{\alpha _{u,k}}} \right\}$, and the transmit BFs, $\left\{ {{{\mathbf{w}}_k}} \right\}$, are highly coupled. Note that for such a challenging optimization problem, it is non-trivial to find the globally optimal solution. In the following, we still invoke the penalty-based method and the SCA method to find a high-quality near-optimal solution.
\subsection{Proposed Solution}
Let us define ${{\mathbf{W}}_k} = {{\mathbf{w}}_k}{\mathbf{w}}_k^H$, which satisfy that ${{\mathbf{W}}_{k}}\succeq 0$ and ${\rm{Rank}}\left( {{{\mathbf{W}}_{k}}} \right)=1,\forall k \in {\mathcal{K}}$. Problem \eqref{cluster based 0} can be reformulated as follows:
\begin{subequations}\label{cluster based 1}
\begin{align}
\mathop {\max }\limits_{\left\{\! {{{\mathbf{W}}_k},{\alpha _{m,k}},{\alpha _{u,k}}} \!\right\},{{\mathbf{R}}_2}}&\!\sum\nolimits_{k = 1}^K\! {{{\log }_2}\!\left(\! {1 \!+ \!\frac{{{\alpha _{u,k}}\!{\rm{Tr}}\left( {{{\mathbf{H}}_{c,k}}{{\mathbf{W}}_k}} \right)}}{{\sum\nolimits_{i \ne k}^K \!{{\rm{Tr}}\left( {{{\mathbf{H}}_{c,k}}{{\mathbf{W}}_i}} \right) \!+ } \sigma _{c,k}^2}}} \!\right)} \\
\label{CT C QoS cluster}{\rm{s.t.}}\;\;&\begin{gathered}
  {\alpha _{m,k}} - {{\overline \gamma }_{m,k}}\frac{{\sum\nolimits_{i \ne k}^K {{\rm{Tr}}\left( {{{\mathbf{H}}_{l,k}}{{\mathbf{W}}_i}} \right) + \sigma _{l,k}^2} }}{{{\rm{Tr}}\left( {{{\mathbf{H}}_{l,k}}{{\mathbf{W}}_k}} \right)}} \hfill \\
    - {{\overline \gamma }_{m,k}}{\alpha _{u,k}}\ge 0,\forall l \in \left\{ {r,c} \right\},k \in {{\mathcal{K}}}, \hfill \\
\end{gathered} \\
\label{rank 1 cluster}&{{\mathbf{W}}_{k}}\succeq 0,{{\mathbf{W}}_{k}}\in {{\mathbb{H}}^N},\forall k \in {\mathcal{K}},\\
\label{rank cluster}&{\rm{Rank}}\left( {{{\mathbf{W}}_{k}}} \right)=1,\forall k \in {\mathcal{K}},\\
\label{constraints cluster 1}&\eqref{cluster radar QoS}-\eqref{power allocation 2},
\end{align}
\end{subequations}
where \eqref{CT C QoS cluster} follows from \eqref{cluster QoS 0} with ${{\mathbf{H}}_{l,k}} \triangleq {{\mathbf{h}}_{l,k}}{\mathbf{h}}_{l,k}^H,\forall l \in \left\{ {r,c} \right\},k \in {\mathcal{K}}$, and ${{\overline \gamma }_{m,k}} = {2^{{{\overline R }_{m,k}}}} - 1,\forall k \in {\mathcal{K}}$. The reformulated problem \eqref{cluster based 1} is a non-convex optimization problem due to the non-convex objective function and the non-convex constraints \eqref{CT C QoS cluster} and \eqref{rank cluster}. Similarly, the key steps of the proposed solution for solving problem \eqref{cluster based 1} are depicted in Fig. \ref{CB}. In the following, we first deal with the non-convex objective function and the constraint \eqref{CT C QoS cluster}.\\
\indent We first introduce some auxiliary variables such that
\begin{align}\label{auxiliary 1}
{\varpi _{c,k}} = \frac{{{\alpha _{u,k}}{\rm{Tr}}\left( {{{\mathbf{H}}_{c,k}}{{\mathbf{W}}_k}} \right)}}{{\sum\nolimits_{i \ne k}^K {{\rm{Tr}}\left( {{{\mathbf{H}}_{c,k}}{{\mathbf{W}}_i}} \right) + } \sigma _{c,k}^2}},\forall k \in {\mathcal{K}},
\end{align}
\begin{align}\label{auxiliary 2}
A_{l,k}^2 = \sum\nolimits_{i \ne k}^K {{\rm{Tr}}\left( {{{\mathbf{H}}_{l,k}}{{\mathbf{W}}_i}} \right) + \sigma _{l,k}^2} ,\forall l \in \left\{ {r,c} \right\},\forall k \in {\mathcal{K}}.
\end{align}
Then, problem \eqref{cluster based 1} can be equivalently rewritten as the following optimization problem:
\begin{subequations}\label{cluster based 2}
\begin{align}
&\mathop {\max }\limits_{\left\{ {{{\mathbf{W}}_k},{\alpha _{m,k}},{\alpha _{u,k}},{\varpi _{c,k}},{A_{c,k}},{A_{r,k}}} \right\},{{\mathbf{R}}_2}} \;\sum\nolimits_{k = 1}^K {{{\log }_2}\left( {1 + {\varpi _{c,k}}} \right)}  \\
\label{CT C QoS cluster 2}{\rm{s.t.}}\;\;&\begin{gathered}
  {\alpha _{m,k}} - {{\overline \gamma }_{m,k}}\frac{{A_{l,k}^2}}{{{\rm{Tr}}\left( {{{\mathbf{H}}_{l,k}}{{\mathbf{W}}_k}} \right)}} \hfill \\
   - {{\overline \gamma }_{m,k}}{\alpha _{u,k}} \ge 0,\forall l \in \left\{ {r,c} \right\},k \in {{\mathcal{K}}}, \hfill \\
\end{gathered} \\
\label{w}&{\varpi _{c,k}} \le \frac{{{\alpha _{u,k}}{\rm{Tr}}\left( {{{\mathbf{H}}_{c,k}}{{\mathbf{W}}_k}} \right)}}{{\sum\nolimits_{i \ne k}^K {{\rm{Tr}}\left( {{{\mathbf{H}}_{c,k}}{{\mathbf{W}}_i}} \right) + } \sigma _{c,k}^2}},\forall k \in {\mathcal{K}},\\
\label{A}&A_{l,k}^2 \ge \sum\nolimits_{i \ne k}^K {{\rm{Tr}}\left( {{{\mathbf{H}}_{l,k}}{{\mathbf{W}}_i}} \right) + \sigma _{l,k}^2} ,\forall l \in \left\{ {r,c} \right\}, k \in {\mathcal{K}},\\
\label{constraints cluster 2}&\eqref{cluster radar QoS}-\eqref{power allocation 2},\eqref{rank 1 cluster},\eqref{rank cluster}.
\end{align}
\end{subequations}
This is because at the optimal solution of \eqref{cluster based 2}, it may be readily verified that constraints \eqref{w} and \eqref{A} will always be satisfied with equality. To demonstrate this, let us assume that at the optimal solution of \eqref{cluster based 2}, we can always increase ${\varpi _{c,k}}$ for ensuring that \eqref{w} is met with strict equality if the constraint \eqref{w} is satisfied with strict inequality. This also increases the value of the objective function. Moreover, if the constraint \eqref{A} is satisfied with strict inequality, we can decrease $A_{l,k}$ for ensuring that \eqref{A} is satisfied with strict equality, without decreasing the value of the objective function at the same time. Therefore, problem \eqref{cluster based 2} is equivalent to problem \eqref{cluster based 1}.\\
\indent For problem \eqref{cluster based 2}, the objective function is concave with respect to ${\varpi _{c,k}}$ and the third term in the LHS of \eqref{CT C QoS cluster 2} is concave jointly with respect to $A_{l,k}$ and ${{\rm{Tr}}\left( {{{\mathbf{H}}_{l,k}}{{\mathbf{W}}_k}} \right)}$. However, the constraints \eqref{w} and \eqref{A} are non-convex with respect to the corresponding optimization variables. To handle the non-convex constraint \eqref{w}, we introduce another auxiliary variable so that
\begin{align}\label{B}
\begin{gathered}
  {\varpi _{c,k}}\left( {\sum\nolimits_{i \ne k}^K {{\rm{Tr}}\left( {{{\mathbf{H}}_{c,k}}{{\mathbf{W}}_i}} \right) + } \sigma _{c,k}^2} \right) \le B_{c,k}^2 \hfill \\
   \le {\alpha _{u,k}}{\rm{Tr}}\left( {{{\mathbf{H}}_{c,k}}{{\mathbf{W}}_k}} \right),\forall k \in {{\mathcal{K}}}. \hfill \\
\end{gathered}
\end{align}
Then, \eqref{w} can be equivalently transformed into the following two constraints:
\begin{figure}[!t]
  \centering
  \includegraphics[width=2.2in]{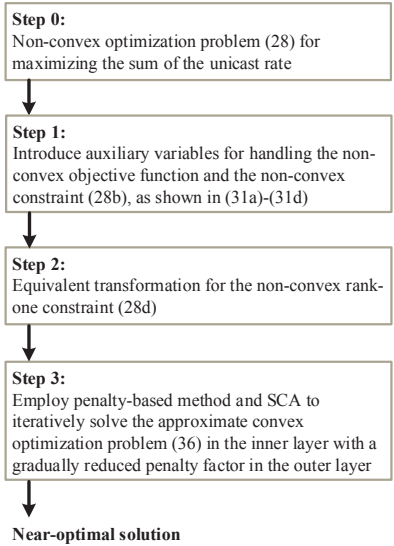}\\
  \caption{Illustration of the key steps in solving problem \eqref{cluster based 1}.}\label{CB}
\end{figure}
\begin{subequations}
\begin{align}\label{EQ 1}
&\sum\nolimits_{i \ne k}^K {{\rm{Tr}}\left( {{{\mathbf{H}}_{c,k}}{{\mathbf{W}}_i}} \right) + } \sigma _{c,k}^2 \le \frac{{B_{c,k}^2}}{{{\varpi _{c,k}}}},\forall k \in {\mathcal{K}}, \\
\label{EQ 2}&\frac{{B_{c,k}^2}}{{{\alpha _{u,k}}}} \le {\rm{Tr}}\left( {{{\mathbf{H}}_{c,k}}{{\mathbf{W}}_k}} \right),\forall k \in {\mathcal{K}}.
\end{align}
\end{subequations}
It can be observed that the constraint \eqref{EQ 1} is non-convex, since the right-hand-side (RHS) is not concave, while the constraint \eqref{EQ 2} is convex. However, the RHS of \eqref{EQ 1} is a convex function joint with respect to ${B_{c,k}}$ and ${{\varpi _{c,k}}}$. For any given feasible points, $\left\{ {B_{c,k}^{n },\varpi _{c,k}^{n }} \right\}$, a lower bound of the RHS of \eqref{EQ 1} is given by
\begin{align}\label{lower bound B}
\begin{gathered}
  \frac{{B_{c,k}^2}}{{{\varpi _{c,k}}}} \ge \frac{{B_{c,k}^{n 2}}}{{\varpi _{c,k}^{n }}} + \frac{{2B_{c,k}^{n }}}{{\varpi _{c,k}^{n }}}\left( {{B_{c,k}} - B_{c,k}^{n }} \right) - \frac{{B_{c,k}^{n 2}}}{{\varpi _{c,k}^{n 2}}}\left( {{\varpi _{c,k}} - \varpi _{c,k}^{n }} \right) \hfill \\
  \;\;\;\;\;\;\;\;=\frac{{2B_{c,k}^{n }}}{{\varpi _{c,k}^{n }}}{B_{c,k}} -  \frac{{B_{c,k}^{n 2}}}{{\varpi _{c,k}^{n 2}}}{\varpi _{c,k}} \triangleq {\Gamma}\left( {{B_{c,k}},{\varpi _{c,k}}} \right),\forall k \in {\mathcal{K}} \hfill \\
\end{gathered}
\end{align}
at the $n$th iteration of SCA by exploiting the first-order Taylor expression. For the non-convex constraint \eqref{A}, a lower bound using the first-order Taylor expression can be obtained as follows:
\begin{align}\label{lower bound A}
A_{l,k}^2 \ge A_{l,k}^{n 2} + 2A_{l,k}^{n }\left( {{A_{l,k}} - A_{l,k}^{n }} \right) \triangleq \Upsilon \left( {{A_{l,k}}} \right),
\end{align}
since its LHS is a convex function with respect to $A_{l,k}$, where ${A_{l,k}^{n }}$ is the given feasible point at the $n$th iteration of the SCA for $ l \in \left\{ {r,c} \right\},k \in {\mathcal{K}}$.\\
\indent As for the remaining non-convex rank-one constraint \eqref{rank cluster}, it can be handled in a same manner as introduced in the previous section. Therefore, by exploiting the penalty-based method as well as the above lower bounds of \eqref{lower bound B} and \eqref{lower bound A}, we have the following optimization problem at the $n$th iteration of the SCA:
\begin{subequations}\label{cluster based 3}
\begin{align}
\mathop {\min }\limits_{\mathcal{X},{\mathbf{R}}_2} & \!- \!\sum\nolimits_{k = 1}^K \!{{{\log }_2}\left( {1 + {\varpi _{c,k}}} \right)}  \!+ \!\frac{1}{{{\eta _2}}}\left( {\sum\nolimits_{k = 1}^K {{{\left\| {{{\mathbf{W}}_k}} \right\|}_*} \!-\! {\overline {\mathbf{W}} _k^n}} } \right)\\
\label{w 3}{\rm{s.t.}}\;\;&\sum\nolimits_{i \ne k}^K {{\rm{Tr}}\left( {{{\mathbf{H}}_{c,k}}{{\mathbf{W}}_i}} \right) + } \sigma _{c,k}^2 \le \Gamma \left( {{B_{c,k}},{\varpi _{c,k}}} \right),\forall k \in {{\mathcal{K}}},\\
\label{A 3}&\Upsilon \left( {{A_{l,k}}} \right) \!\ge\! \sum\nolimits_{i \ne k}^K \!{{\rm{Tr}}\left( {{{\mathbf{H}}_{l,k}}{{\mathbf{W}}_i}} \right) \!+\! \sigma _{l,k}^2} ,\forall l \!\in \!\left\{ {r,c} \right\}, \!k\! \in\! {{\mathcal{K}}},\\
\label{constraints cluster 3}&\eqref{cluster radar QoS}-\eqref{power allocation 2},\eqref{rank 1 cluster},\eqref{CT C QoS cluster 2},\eqref{EQ 2},
\end{align}
\end{subequations}
where $\overline {\mathbf{W}} _k^n \triangleq  - {\left\| {{\mathbf{W}}_k^n} \right\|_2} - {\rm{Tr}}\left[ {{{\mathbf{v}}_{\max }}\left( {{\mathbf{W}}_k^n} \right){\mathbf{v}}_{\max }^H\left( {{\mathbf{W}}_k^n} \right)\left( {{{\mathbf{W}}_k} - {\mathbf{W}}_k^n} \right)} \right],\forall k \in {\mathcal{K}}$ and ${{\mathcal{X}}} \triangleq \left\{ {\left\{ {{{\mathbf{W}}_k},{\alpha _{m,k}},{\alpha _{u,k}},{\varpi _{c,k}},{A_{c,k}},{A_{r,k}},{B_{c,k}}} \right.} \right\}$. Now, for any given penalty factor, $\eta_2>0$, it may be readily shown that problem \eqref{cluster based 3} is a convex optimization problem, which can be efficiently solved using CVX~\cite{cvx}. In order to obtain feasible rank-one matrix solutions of high performance, we still develop a double-layer penalty-based algorithm. In the inner layer, problem \eqref{cluster based 3} is iteratively solved by employing SCA for a given $\eta_2$. In the outer layer, the value of $\eta_2$ is gradually decreased for ensuring that the matrix solutions obtained become of rank-one. Similarly, the algorithm terminates, when the equality constraints are satisfied with the predefined accuracy, yielding the following condition:
\begin{align}\label{termination 2}
\max \left\{ {{{\left\| {{{\mathbf{W}}_k}} \right\|}_*} - {{\left\| {{{\mathbf{W}}_k}} \right\|}_2},\forall k \in {{\mathcal{K}}}} \right\} \le {{\epsilon}_o}.
\end{align}
\indent The details of the double-layer penalty-based procedure developed for solving problem \eqref{cluster based 0} are summarized in \textbf{Algorithm 2}, which is guaranteed to converge to a stationary solution of the original problem \eqref{cluster based 0}~\cite{SCA}. The main computational complexity of \textbf{Algorithm 2} arises from iteratively solving problem \eqref{cluster based 0}. If the inner-point method of~\cite{convex} is employed, the complexity of solving problem \eqref{cluster based 0} is on the order of ${\mathcal{O}}\left( {K{N^{3.5}} + {{\left( {6K} \right)}^{3.5}}} \right)$, where ${6K}$ denotes the number of scalar optimization variables. As a result, the total computational complexity of \textbf{Algorithm 2} is given by ${\mathcal{O}}\left[ {I_o^2I_i^2\left( {K{N^{3.5}} + {{\left( {6K} \right)}^{3.5}}} \right)} \right]$, where ${I_i^2}$ and ${I_o^2}$ denotes the number of inner and outer iterations required for convergence of \textbf{Algorithm 2}.
\begin{algorithm}[!t]\label{method2}
\caption{Proposed double-layer penalty-based algorithm for solving the joint BF design and power allocation problem \eqref{cluster based 0}}
\begin{algorithmic}[1]
\STATE {Initialize feasible points $\left\{ {{\mathbf{W}}_k^0,\alpha _{m,k}^0,\alpha _{u,k}^0} \right\}$ and the penalty factor $\eta_2$.}
\STATE {\bf repeat: outer layer}
\STATE \quad Set iteration index $n=0$ for inner layer.
\STATE \quad {\bf repeat: inner layer}
\STATE \quad\quad Calculate the current value of $\left\{ {\varpi _{c,k}^n,A_{c,k}^n,A_{r,k}^n,B_{c,k}^n} \right\}$ using \eqref{auxiliary 1}, \eqref{auxiliary 2}, and \eqref{B}.
\STATE \quad\quad For given feasible points, solve the convex problem \eqref{cluster based 3} and the solutions obtained are denoted by $\left\{ {{\mathbf{W}}_k^*,\alpha _{m,k}^*,\alpha _{u,k}^*} \right\}$.
\STATE \quad\quad Update $\left\{ {{\mathbf{W}}_k^{n + 1},\alpha _{m,k}^{n + 1},\alpha _{u,k}^{n + 1}} \right\}$ by the obtained optimal solutions and $n=n+1$.
\STATE \quad {\bf until} the fractional reduction of the objective function value falls below a predefined threshold $\epsilon_i >0$.
\STATE \quad Update $\left\{ {{\mathbf{W}}_k^0,\alpha _{m,k}^0,\alpha _{u,k}^0} \right\}$ by the currently obtained optimal solutions.
\STATE \quad Update $\eta_1  = \varepsilon \eta_1$.
\STATE {\bf until} the constraint violation falls below a maximum tolerable threshold $\epsilon_o >0$.
\end{algorithmic}
\end{algorithm}
\section{Numerical Results}
In this section, we provide numerical results obtained by Monte Carlo simulations for characterizing the proposed NOMA-aided joint Rad-MU-Com frameworks. In particular, we assume that the DFRC BS employs a uniform linear array (ULA) with half-wavelength spacing between adjacent antennas. The channel between the BS and the R-user is assumed to have pure line-of-sight (LoS) associated with the path loss of ${L_R} = {L_0} + 20{\log _{10}}{d_R}$, while between the BS and C-user it is assumed to obey the Rayleigh channel model with the path loss of ${L_C} = {L_0} + 30{\log _{10}}{d_C}$~\cite{protocol,security}, where ${L_0}$ is the path loss at the reference distance $d=1$ meter (m), and ${d_R}$ and ${d_C}$ represents the distance from the BS to the R-user and to the C-user, respectively. The parameters adopted in simulations are set as follows: ${L_0}=40$ dB, ${d_R}=1000$ m, and ${d_C}=100$ m. The noise power in the receiver of users is assumed to be the same, which is set to ${\sigma ^2}=-100$ dBm. The transmit-signal-to-noise-ratio (SNR)\footnote{Using the transmit-SNR is unconventional, because it is given by the ratio of the transmit power and the receiver noise, which are quantities measured at different points. This quantity is however beneficial for our joint Rad-Com problem, where the optimum transmit power is assigned to each user for satisfying their individual rate requirements under the idealized simplifying assumption that they have perfect capacity-achieving receivers relying on powerful capacity-achieving channel codes. This is because the optimization problem of our specific system was formulated for maximizing the unicast performance at a given transmit power, while satisfying specific constraints imposed both on the multicast rate and on the radar beam pattern.} is considered in the simulations, which is given by ${\gamma _p} = \frac{{{P_{\max }}}}{{{\sigma ^2}}}$. The initial penalty factors of \textbf{Algorithms 1} and \textbf{2} are set to ${\eta _1} = {\eta _2} = {10^4}$, the convergence threshold of the inner layer is set to $\epsilon_i = 10^{-2}$, and the algorithm's termination threshold of the equality constraints is set to $\epsilon_o = 10^{-5}$. The numerical results were obtained by averaging over 200 channel realizations. \\
\indent To obtain the optimal solutions (i.e., ${\mathbf{R}_0^*}$ and ${\delta ^*}$) and the performance benchmark (i.e., $\Delta \left( {{\mathbf{R}}_0^*,{\delta ^*}} \right)$) of the radar-only system in problem \eqref{ideal beam pattern design}, the desired beam pattern, ${P^*}\left( {{\theta _m}}\right)$, is defined as follows:
\begin{align}\label{P}
{P^*}\left( {{\theta _m}} \right) = \left\{ \begin{gathered}
  1,\;\;{\theta _m} \in \left[ {{\overline \theta _k} - \frac{\Delta }{2},{\overline \theta _k} + \frac{\Delta }{2}} \right],\forall k \in {\mathcal{K}}, \hfill \\
  0,\;\;{\rm{otherwise}}, \hfill \\
\end{gathered}  \right.
\end{align}
where $\left\{ {{\overline \theta _k},\forall k \in {\mathcal{K}}} \right\}$ denotes the actual true angles to be detected, which are determined by the location of R-users, and $\Delta $ denotes the width of the desired beam, which is set to ${10^ \circ }$ in the simulations.
\subsection{BB NOMA-Aided Joint Rad-MU-Com System}
In the BB NOMA-aided joint Rad-MU-Com system (also referred to as ``BB NOMA+Rad-MU-Com''), we assume that the R-user is located at the angle of ${0^ \circ }$. \subsubsection{Benchmark Schemes} For performance comparison, we consider the following two benchmark schemes, which have been discussed in \textbf{Remark 2}.
\begin{itemize}
  \item \textbf{TDMA-based joint Rad-MU-Com system (also referred to as ``TDMA+Rad-MU-Com'')}: In this scheme, the MIMO DFRC BS successively transmits the multicast and unicast messages to the R- and C-users over two time slots employing one BF{\footnote{In contrast to the communication-only systems, where TDMA can successively employ two different BFs to deliver the multicast and unicast messages, it is practically relevant to assume that the joint Rad-MU-Com system has to employ a common BF for successively conveying different messages, which guarantee that the corresponding beam pattern remains unchanged for radar detection.}}, which is also used for detecting the R-user. Accordingly, for TDMA+Rad-MU-Com, the achievable rate of the multicast signal at the R- and C-users is given by
      \begin{align}\label{TDMA CT rate CT}
      R_{m \to l}^{{\rm{TDMA}}} = \frac{1}{2}{\log _2}\left( {1 + \frac{{{{\left| {{\mathbf{h}}_l^H{\mathbf{w}}} \right|}^2}}}{{\sigma _l^2}}} \right),\forall l \in \left\{ {r,c} \right\}.
      \end{align}
      The corresponding rate of the unicast signal at the C-user is
      \begin{align}\label{TDMA C rate C}
      R_u^{{\rm{TDMA}}} = \frac{1}{2}{\log _2}\left( {1 + \frac{{{{\left| {{\mathbf{h}}_c^H{\mathbf{w}}} \right|}^2}}}{{\sigma _c^2}}} \right).
      \end{align}
      The problem of maximizing $R_u^{{\rm{TDMA}}}$ can be solved by using \textbf{Algorithm 1}, but without the inter-signal interference term.
  \item \textbf{CBF-No-SIC-based joint Rad-MU-Com system (also referred to as ``CBF-No-SIC+Rad-MU-Com'')}: In this scheme, the MIMO DFRC BS simultaneously transmits the multicast and unicast messages to the R- and C-users employing two different BFs, which are also jointly used to detect the R-user. All users will directly detect their intended signals by treating others as interference without the assistance of SIC. Therefore, the rate achieved for the unicast signal at the C-user is given by
      \begin{align}\label{SDMA C rate C}
      {R_{u}^{{\rm{CBF-No-SIC}}}} = {\log _2}\left( {1 + \frac{{{{\left| {{\mathbf{h}}_c^H{{\mathbf{w}}_u}} \right|}^2}}}{{{{\left| {{\mathbf{h}}_c^H{{\mathbf{w}}_{m}}} \right|}^2} + \sigma _c^2}}} \right).
      \end{align}
      Note that, for CBF-No-SIC+Rad-MU-Com, the expressions of the rate achieved for the multicast signal at the R- and C-users are the same as \eqref{CT rate C} and \eqref{CT rate T}. The resultant optimization problem of maximizing ${R_{u}^{{\rm{CBF-No-SIC}}}}$ can be solved following a similar process to that of \textbf{Algorithm 2} to deal with the interference term in \eqref{SDMA C rate C}.
\end{itemize}
Although the employment of a single BF in the TDMA+Rad-MU-Com system limits the DoFs compared to that of NOMA and CBF-No-SIC, the advantage is that TDMA supports interference-free transmission for delivering both types of messages. The performance obtained by the proposed NOMA scheme and the two benchmark schemes will be compared in the following.
\subsubsection{Unicast Rate Versus ${{\overline \gamma}_{b}}$} 
\begin{figure}[!t]
  \centering
  \includegraphics[width=2.9in]{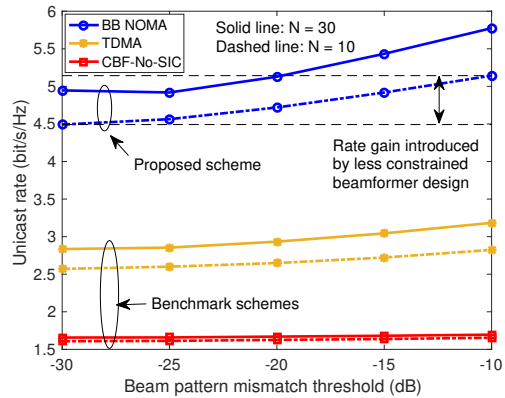}\\
  \caption{The unicast rate versus ${{\overline \gamma}_{b}}$ for ${\gamma _p}=110$ dB and ${\overline R}_{m}=0.5$ bit/s/Hz.}\label{Rvrb}
\end{figure}

In Fig. \ref{Rvrb}, we investigate the unicast rate, $R_u$, achieved versus the maximum tolerable beam pattern mismatch, ${{\overline \gamma}_{b}}$. We set ${\gamma _p}=110$ dB (i.e., ${P_{\max}}=10$ dBm) and ${\overline R}_{m}=0.5$ bit/s/Hz. As seen in Fig. \ref{Rvrb}, the unicast rate obtained by all schemes increases as ${{\overline \gamma}_{b}}$ increases. This is indeed expected, since lager ${{\overline \gamma}_{b}}$ values impose looser constraints on the BF design, which provides higher DoFs, hence enhancing the unicast performance. Moreover, a higher $N$ leads to a higher unicast rate due to the enhanced array gain and spatial DoFs. By comparing the three Rad-MU-Com schemes presented, it may be observed that the proposed BB NOMA+Rad-MU-Com scheme achieves the best performance. This is because on the one hand, employing SIC in NOMA mitigates the inter-signal interference compared to the CBF-No-SIC+Rad-MU-Com scheme, thus improving the unicast performance achieved at the C-user. On the other hand, the power-domain resource sharing and the employment of two different BFs allows NOMA to achieve a higher performance than the TDMA+Rad-MU-Com scheme. Moreover, despite employing a single BF, the TDMA+Rad-MU-Com scheme can deliver both types of messages in an interference-free manner, while carrying out radar detection. Therefore, the TDMA+Rad-MU-Com scheme outperforms the CBF-No-SIC+Rad-MU-Com scheme, whose performance is significantly degraded by the inter-signal interference. It can also be observed that the performance gain obtained by NOMA is more noticeable when ${{\overline \gamma}_{b}}$ increases. The above results verify the efficiency of the proposed BB NOMA+Rad-MU-Com framework.
\subsubsection{Unicast Rate Versus ${\overline R}_{m}$} 
\begin{figure}[!t]
  \centering
  \includegraphics[width=2.9in]{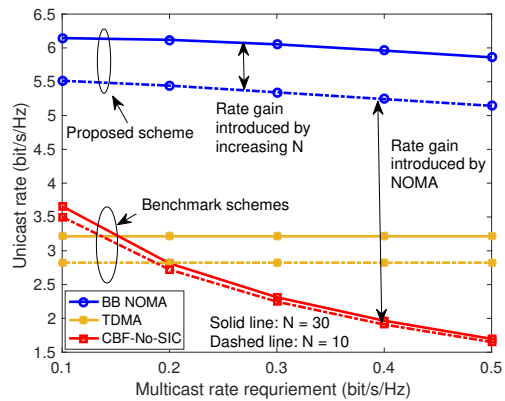}\\
  \caption{The unicast rate versus ${{\overline R}_{m}}$ for ${\gamma _p}=110$ dB and ${{\overline \gamma}_{b}}=-10$ dB.}\label{Rvrt}
\end{figure}

In Fig. \ref{Rvrt}, we investigate the unicast rate, $R_u$, achieved versus the rate requirement of multicast, ${\overline R}_{m}$. We set ${\gamma _p}=110$ dB and ${{\overline \gamma}_{b}}=-10$ dB. As seen from Fig. \ref{Rvrt}, NOMA achieves the best performance. Moreover, the unicast rate obtained by NOMA and CBF-No-SIC decreases as ${\overline R}_{m}$ increases. This is because a higher multicast rate  requires more transmit power to be allocated, thus degrading the unicast rate achieved.
Without the mitigating assistance of SIC on the inter-signal interference, the degradation of the unicast rate in the CBF-No-SIC+Rad-MU-Com scheme is more significant than that in the proposed BB NOMA+Rad-MU-Com scheme. Furthermore, the unicast rate obtained by the TDMA+Rad-MU-Com scheme remains almost unchanged. The reason for this trend is as follows. Recall the fact that the TDMA+Rad-MU-Com scheme employs a single common BF for successively delivering the two kinds of messages, where the rate expressions of unicast and multicast at the C-user are the same, see \eqref{TDMA CT rate CT} and \eqref{TDMA C rate C}. Therefore, when the rate requirement of multicast is lower than the unicast rate, the impact of ${\overline R}_{m}$ on the unicast rate becomes negligible, since the multicast rate requirement is automatically satisfied as long as $R_u$ is higher than ${\overline R}_{m}$. This also reveals the inefficiency of the fixed resource allocation in TDMA. Additionally, we can observe that the CBF-No-SIC+Rad-MU-Com scheme outperforms the TDMA+Rad-MU-Com scheme when ${\overline R}_{m}$ is low, but its performance erodes worse when ${\overline R}_{m}$ increases. This is because, for smaller ${\overline R}_{m}$, the unicast transmission of CBF-No-SIC becomes less contaminated by the interference from the multicast signal, thus achieving a higher unicast performance than TDMA due to its full-time transmission. However, when ${\overline R}_{m}$ becomes stricter, the unicast performance of CBF-No-SIC is significantly degraded by the interference caused by the multicast signal. In this case, the interference-free TDMA outperforms CBF-No-SIC.
\subsubsection{Unicast Rate Versus ${\gamma _p}$} 
\begin{figure}[!t]
  \centering
  \includegraphics[width=2.9in]{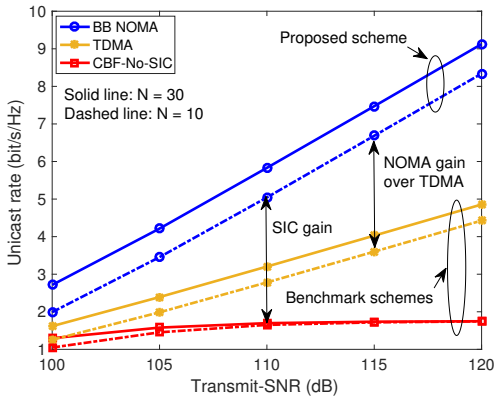}\\
  \caption{The unicast rate versus ${\gamma _p}$ for ${{\overline R}_{m}}=0.5$ bit/s/Hz and ${{\overline \gamma}_{b}}=-10$ dB.}\label{RvP}
\end{figure}

In Fig. \ref{RvP}, we present the unicast rate, $R_u$, achieved versus the transmit-SNR, ${\gamma _p}$. We set ${\overline R}_{m}=0.5$ bit/s/Hz and ${{\overline \gamma}_{b}}=-10$ dB. It can be observed that the unicast rate of all schemes increases upon increasing ${\gamma _p}$. However, in contrast to both NOMA and TDMA, the rate enhancement of CBF-No-SIC attained upon increasing ${\gamma _p}$ becomes negligible and the unicast rate is seen to be bounded by a certain value. This is because when the inter-signal interference is not mitigated, CBF-No-SIC becomes interference-limited, when the transmit power is high. Moreover, it can also be seen from Fig. \ref{RvP} that the rate enhancement attained by NOMA upon increasing ${\gamma _p}$ is more significant than for TDMA, since NOMA benefits from a flexible resource allocation scheme.
\subsubsection{Beam Pattern of a Signal R-User and a Single C-User} 
\begin{figure}[!t]
  \centering
  \includegraphics[width=2.9in]{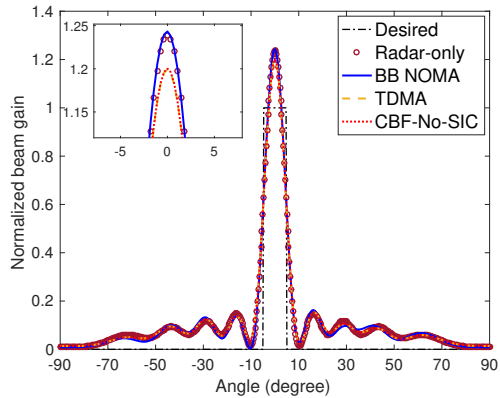}\\
  \caption{The transmit beam pattern obtained by different schemes in the considered Rad-MU-Com system with one R-user and one C-user, where $N=10$, ${\gamma _p}=105$ dB, ${\overline R}_{m}=0.5$ bit/s/Hz, and ${{\overline \gamma}_{b}}=-10$ dB.}\label{Beampattern_BB}
\end{figure}

In Fig. \ref{Beampattern_BB}, we plot the transmit beam pattern obtained by the three schemes for one random channel realization. We set $N=10$, ${\gamma _p}=105$ dB, ${\overline R}_{m}=0.5$ bit/s/Hz, and ${{\overline \gamma}_{b}}=-20$ dB. In particular, the desired beam pattern is obtained according to \eqref{P} and the beam pattern of the radar-only system is obtained by solving problem \eqref{ideal beam pattern design}. As illustrated in Fig. \ref{Beampattern_BB}, the beam pattern obtained by NOMA closely approaches that of the radar-only system, while the beam pattern mismatch of TDMA and CBF-No-SIC becomes more noticeable. Observe in Figs. \ref{Rvrb}-\ref{RvP} that given the same accuracy requirement of the radar beam pattern, the proposed BB NOMA+Rad-MU-Com scheme achieves higher communication performance than the other benchmark schemes. The above results also confirm the effectiveness of the proposed BB NOMA+Rad-MU-Com framework.
\subsection{CB NOMA-Aided Joint MIMO Rad-MU-Com System}
In the CB NOMA-aided joint Rad-MU-Com system (also referred to as ``CB NOMA+Rad-MU-Com''), we consider the case of $K=3$ R\&C pairs\footnote{In this work, the C- and R-users are randomly paired. Note that more sophisticated user pairing strategies can be developed for further improving the performance of the joint Rad-MU-Com system considered, which is beyond the scope of this paper and constitutes an interesting topic for future work.}, where the 3 R-users are assumed to be located at the angles of $\left[ { - {{60}^ \circ },{0^ \circ },{{60}^ \circ }} \right]$. Without loss of generality, the multicast rate requirements of each pair are assumed to be the same, i.e., ${\overline R _{m,k}} = {\overline R _{m,0}},\forall k \in {\mathcal{K}}$.
\subsubsection{Benchmark Scheme} We consider the TDMA+Rad-MU-Com as our benchmark scheme\footnote{As it is impossible for CBF-No-SIC to employ one common BF to deliver two different messages, only the TDMA-based benchmark scheme is considered for the system for multiple R\&C pairs.}. In this case, the DFRC BS successively transmits the unicast and multicast messages employing different BFs intended for each R\&C pair. Accordingly, the communication rate attained for the multicast signal of the R- and C-users in the $k$th pair is given by
\begin{align}\label{TDMA CT rate CT CC}
R_{m \to l,k}^{{\rm{TDMA}}} = \frac{1}{2}{\log _2}\left( {1 + \frac{{{{\left| {{\mathbf{h}}_{l,k}^H{{\mathbf{w}}_k}} \right|}^2}}}{{\sum\nolimits_{i \ne k}^K {{{\left| {{\mathbf{h}}_{l,k}^H{{\mathbf{w}}_i}} \right|}^2}}  + \sigma _{l,k}^2}}} \right),\forall l \in \left\{ {r,c} \right\}.
\end{align}
Then, the communication rate achieved for the unicast signal at the C-user of the $k$th pair becomes:
\begin{align}\label{TDMA C rate C CC}
R_{u,k}^{{\rm{TDMA}}} = \frac{1}{2}{\log _2}\left( {1 + \frac{{{{\left| {{\mathbf{h}}_{c,k}^H{{\mathbf{w}}_k}} \right|}^2}}}{{\sum\nolimits_{i \ne k}^K {{{\left| {{\mathbf{h}}_{c,k}^H{{\mathbf{w}}_i}} \right|}^2}}  + \sigma _{c,k}^2}}} \right).
\end{align}
The resultant optimization problem of maximizing $\sum\nolimits_{k = 1}^K {R_{u,k}^{{\rm{TDMA}}}}$ can be solved by \textbf{Algorithm 2}, but without considering the power allocation.
\subsubsection{Sum of Unicast Rate Versus ${{\overline \gamma}_{b}}$} 
\begin{figure}[!t]
  \centering
  \includegraphics[width=2.9in]{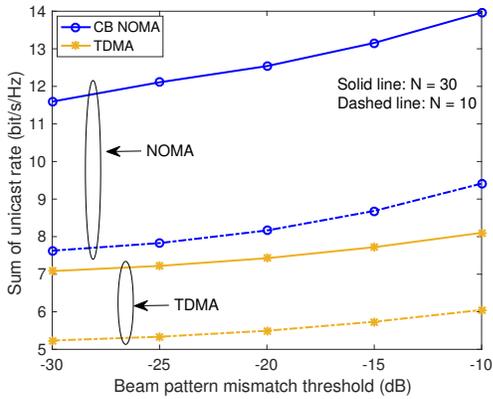}\\
  \caption{The sum of the unicast rate versus ${{\overline \gamma}_{b}}$ for ${\gamma _p}=110$ dB and ${\overline R}_{m}=0.5$ bit/s/Hz.}\label{SRvrb}
\end{figure}

In Fig. \ref{SRvrb}, we investigate the sum of the unicast rate achieved versus the maximum tolerable radar beam pattern mismatch. We set ${\gamma _p}=110$ dB and ${\overline R _{m,0}}=0.5$ bit/s/Hz. Observe that the proposed CB NOMA+Rad-MU-Com scheme outperforms the TDMA+Rad-MU-Com scheme and the sum rate gain of CB NOMA over TDMA becomes more pronounced, when ${{\overline \gamma}_{b}}$ increases. For achieving the same unicast performance, the proposed CB NOMA+Rad-MU-Com scheme requires less number of transmit and can satisfy a stricter beam pattern mismatch constraint, as compared to the TDMA scheme. In other words, NOMA can well intergrade both functions of communication and radar detection.
\begin{figure}[!t]
  \centering
  \includegraphics[width=2.9in]{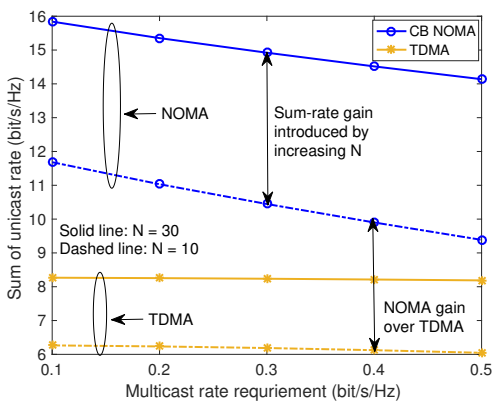}\\
  \caption{The sum of the unicast rate versus ${\overline R _{m,0}}$ for ${\gamma _p}=110$ dB and ${{\overline \gamma}_{b}}=-10$ dB.}\label{SRvrt}
\end{figure}
\subsubsection{Sum of Unicast Rate Versus ${\overline R _{m,0}}$} In Fig. \ref{SRvrt}, we study the sum of the unicast rate versus the multicast rate requirements of each pair. We set ${\gamma _p}=110$ dB and ${{\overline \gamma}_{b}}=-10$ dB. We can observe that the proposed CB NOMA+Rad-MU-Com scheme outperforms the TDMA+Rad-MU-Com scheme, especially when ${\overline R _{m,0}}$ is small. This reveals that the proposed CB NOMA+Rad-MU-Com scheme is more suitable for scenarios having heterogenous rate requirements. Observe from Fig. \ref{SRvrb} that the sum rate gain attained by increasing $N$ for CB NOMA is more significant than that for TDMA, which means that the proposed CB NOMA+Rad-MU-Com scheme can better exploit the spatial DoFs than TDMA.
\subsubsection{Beam Pattern of Multiple R-Users and Multiple C-Users} 
\begin{figure}[!t]
  \centering
  \includegraphics[width=2.9in]{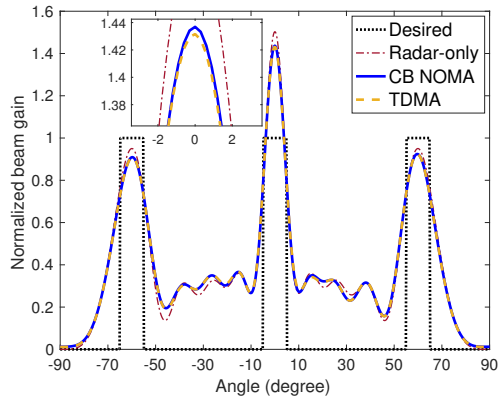}\\
  \caption{The transmit beam pattern obtained by different schemes in the joint Rad-MU-Com system with $K=3$ R\&C pairs, where $N=10$, ${\gamma _p}=110$ dB, ${\overline R}_{m,0}=0.5$ bit/s/Hz, and ${{\overline \gamma}_{b}}=-20$ dB.}\label{Beampattern_CB}
\end{figure}

In Fig. \ref{Beampattern_CB}, we present the beam pattern obtained for the Rad-MU-Com system considered having $K=3$ R\&C pairs for a random channel realization. We set $N=10$, ${\gamma _p}=110$ dB, ${\overline R}_{m,0}=0.5$ bit/s/Hz, and ${{\overline \gamma}_{b}}=-20$ dB. Observe that both the beam patterns obtained by CB NOMA and TDMA approach the beam pattern of the radar-only system upon detecting the 3 R-users. However, the communication performance achieved by NOMA is significantly higher than by TDMA, as seen in Figs. \ref{SRvrb} and \ref{SRvrt}. The above results verify that the proposed scheme is eminently suitable for mixed multicast-unicast transmission while well achieving a high-quality beam pattern for multiple target detection in radar.
\begin{figure}[!t]
  \centering
  \includegraphics[width=2.9in]{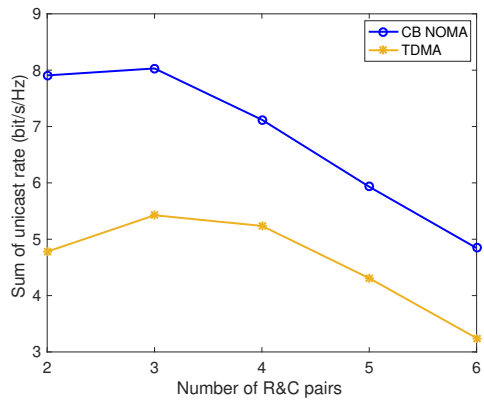}\\
  \caption{The sum of the unicast rate versus $K$, where $N=10$, ${\gamma _p}=110$ dB, ${\overline R}_{m,0}=0.5$ bit/s/Hz, and ${{\overline \gamma}_{b}}=-20$ dB.}\label{SRvK}
\end{figure}
\subsubsection{Sum of Unicast Rate Versus $K$} In Fig. \ref{SRvK}, we further investigate the sum of the unicast rate versus the number of R\&C pairs. We set $N=10$, ${\gamma _p}=110$ dB, ${\overline R}_{m,0}=0.5$ bit/s/Hz, and ${{\overline \gamma}_{b}}=-20$ dB. We consider the cases of $K=2, 3, 4, 5, 6$ with the R-users located at the angle of $\left[ { - {{60}^ \circ },{{60}^ \circ }} \right]$, $\left[ { - {{60}^ \circ },{0^ \circ },{{60}^ \circ }} \right]$, $\left[ { - {{60}^ \circ },- {{30}^ \circ },{{30}^ \circ },{{60}^ \circ }} \right]$, $\left[ { - {{60}^ \circ },- {{30}^ \circ },{0^ \circ },{{30}^ \circ },{{60}^ \circ }} \right]$, $\left[ { - {{60}^ \circ },- {{40}^ \circ },- {{20}^ \circ }, {{20}^ \circ },{{40}^ \circ },{{60}^ \circ }} \right]$. It can be observed that the sum rate of the two schemes first increases and then decreases with the increase of $K$. The reason for this trend is as follows. When $K$ is small, increasing the number of R\&C pairs (i.e., from 2 to 3) provides more spatial DoFs to be exploited for maximizing the sum rate. However, when $K$ becomes large, high transmit power has to be allocated to satisfy the multicast communication requirement of each R\&C pair. This, in turn, reduces the transit power available for unicast communication, thus leading to a degraded sum of the unicast rate performance.
\section{Conclusions}
A novel NOMA-aided joint Rad-MU-Com concept has been proposed, where a MIMO DFRC BS transmits superimposed multicast and unicast messages to the R- and C-users, while detecting the R-user target. The BB NOMA and CB NOMA-aided joint Rad-MU-Com frameworks were proposed for the systems supporting a single and multiple pairs of R- and C-users, respectively. For each framework, tailor-made BF optimization problems were formulated for enhancing the unicast performance, while satisfying both the multicast rate and the radar beam pattern requirements. To solve the resultant non-convex optimization problems, penalty-based iterative algorithms were developed to find a near-optimal solution. The numerical results obtained revealed that a higher unicast performance can be achieved by the proposed NOMA-aided joint Rad-MU-Com schemes than by the benchmark schemes.
\bibliographystyle{IEEEtran}
\bibliography{mybib}

\end{document}